\documentstyle[preprint,aps,floats,psfig]{revtex}
\begin{document}

\title  {Local density of states for the corner geometry interface of
 $d$-wave superconductors, within the extended Hubbard model
 }
\author{N. Stefanakis}
\address{ Department of Physics, University of Crete,
        P.O. Box 2208, GR-71003, Heraklion, Crete, Greece}
\date{\today}
\maketitle

\begin{abstract}
The spatial variations of the order parameter,
and the local density of
states (LDOS) on the corner of $s$-wave or 
$d_{x^2-y^2}$-wave superconductors, as
well as in superconductor-insulator-normal metal interfaces,  are
calculated self consistently using the Bogoliubov-deGennes formalism within
the two dimensional extended Hubbard model.
The exact diagonalization method is used. 
Due to the
suppression of the dominant $d$-wave order parameter, 
the extended $s$-wave order parameter is induced near the surface,
that alternates its sign for the topmost sites
at adjacent edges of the lattice and decays to zero in the bulk.
The presence of
surface roughness results into the appearance of the
zero band conduction peak (ZBCP) near the corner surface which lacks from the
predictions of the quasiclassical theory.
\end{abstract}

\section{Introduction}

The determination of the order parameter symmetry has become one of the 
main aspects in the research on high temperature superconductors
\cite{scalapino,vanh}.
Tunneling conductance experiments report the existence of a zero band 
conduction peak (ZBCP) \cite{green,lesueur}. 
The origin of the experimental 
ZBCP is explained 
in the context of zero energy states (ZES) \cite{hu} formed near 
the $[110]$ surfaces of 
$d$-wave superconductors \cite{tanaka,zhu}. 
These ZES do not appear for $s$-wave 
superconductors or near the $[100]$ surface of $d$-wave superconductors 
and are one of the features that characterize the $d$-wave 
superconductors.

The quasiclassical theory of superconductivity has been used to calculate 
the tunneling conductance in interfaces of unconventional superconductors 
with normal metals or ferromagnets \cite{stefan1,stefan2,stefan3}. 
In the quasiclassical approximation  
the quasiparticles move in classical trajectories (external degrees 
of freedom) with internal degrees of freedom which are the spin and 
the particle-hole degrees of freedom. 
The orientation dependence 
of the spectra as well as the $V$ line shape of the conductance 
curve are explained by the formation of bound states close to the 
interface due to the sign change of the pair potential that the 
transmitted quasiparticles experience.
 
Moreover the concept of phase shift by $\pi$ of the order parameter 
in orthogonal directions in $k$-space, which is equivalent to the 
sign change of the Josephson critical current, can be observed in corner 
junctions of anisotropic superconductors with conventional 
$s$-wave superconductors as a dip of the Fraunhofer pattern
at zero magnetic field \cite{vanh}. 
It is an indication of $d$-wave symmetry 
of the order parameter. The spontaneous flux modulation with surface
orientation in such junctions has been calculated and can be used 
to distinguish the subdominant components $s$ or $d_{xy}$ 
that are induced at regions 
where the $d$-wave order parameter is suppressed.

The extended Hubbard model has been used to study single vortex structure 
\cite{kallin1,wang}, time reversal symmetry breaking across twin boundaries
\cite{kallin3} or near surfaces \cite{zhu}, 
the effect of disorder \cite{kallin2}, 
and the effect of surface roughness \cite{tanaka}. 
In this paper the Bogoliubov-deGennes equations are solved in a 
two dimensional square lattice within the context 
of an extended Hubbard model. The spatial variation of order parameter, 
and the local density of states, which in the limit of low transparency 
barrier converges to the tunneling conductance, 
are calculated for various types of 
surfaces and interfaces e.g. a corner surface, the interface of $d$-wave or 
$s$-wave superconductor along the $[110]$ direction  
with normal metals. 
The evolution of the local density of states (LDOS) is studied as a function of 
the distance from the surface.

It is seen that the extended $s$-wave order parameter is induced due to the 
suppression of the dominant $d$-wave order parameter which 
alternates its sign for the topmost sites 
at adjacent edges of the lattice and decays to zero in the bulk. 
The LDOS is symmetric when $\mu=0$ and it
becomes asymmetric when $\mu$ deviates from zero due to the
breakdown of the electron-hole symmetry.
We also investigate the effect of the surface roughness near the 
corner. 
In general surface roughness which in real samples
is of atomic length scale modifies
the properties of the quasiparticles since the coherence length of the
cuprates is much smaller than the conventional $s$-wave superconductors.
Our model treats the quasiparticle properties on atomic length 
scale and goes beyond the quasiclassical approximation.
The presence of 
surface roughness results into the appearance of the 
ZBCP near the corner surface which lacks from the 
predictions of the quasiclassical theory. 
 
The rest of the paper is organized as following. In Sec. II we
develop the model and discuss the formalism. In Sec. III we present
the results for the corner of superconductor. In Sec. IV we present
the $s-I-n$, $d-I-n$ interfaces. 
In Sec. V, the effect of the surface roughness
is considered.
Finally, summary and discussions are presented in the last section.

\section{BdG equations, within the Hubbard model} 

The Hamiltonian for the extended Hubbard model on a two dimensional square
lattice is
\begin{eqnarray}
H & = & -t\sum_{<i,j>\sigma}c_{i\sigma}^{\dagger}c_{j\sigma} 
+\mu \sum_{i\sigma} n_{i\sigma}+\sum_{i\sigma} \mu_i^In_{i\sigma} \nonumber \\
  & + & V_0\sum_{i} n_{i\uparrow} n_{i\downarrow}
+\frac{V_1}{2}\sum_{<ij>\sigma\sigma^{'}} n_{i\sigma} n_{j\sigma^{'}}\, 
,~~~\label{bdgH}
\end{eqnarray}
where $i,j$ are sites indices and the angle brackets indicate that the 
hopping is only to nearest neighbors, 
$n_{i\sigma}=c_{i\sigma}^{\dagger}c_{i\sigma}$ is the electron number 
operator in site $i$, $\mu$ is the chemical potential. $V_0$, $V_1$  
are on site and nearest-neighbor interaction strength. Negative values 
of $V_0$ and $V_1$ mean attractive interaction and positive values mean 
repulsive interaction. When $V_1<0$ 
the pairing interaction gives rise to the $d$-wave 
superconductivity in a restricted parameter regime \cite{wang}. 
To simulate the effect of the depletion of the carrier density 
at the surface or impurities the site-dependent 
impurity potential $\mu^I(r_i)$ is set to a sufficiently large  value 
at the surface sites. This prohibit the electron tunneling over these
sites. Within the mean field approximation Eq. (\ref{bdgH}) is reduced  to 
the Bogoliubov deGennes equations \cite{gennes}:
\begin{equation} 
\left(
\begin{array}{ll}
  \hat{\xi} & \hat{\Delta} \\
  \hat{\Delta}^{\ast} & -\hat{\xi} 
\end{array}
\right)
\left(
\begin{array}{ll}
  u_n(r_i) \\
  v_n(r_i) 
\end{array}
\right)
=\epsilon_n
\left(
\begin{array}{ll}
  u_n(r_i) \\
  v_n(r_i) 
\end{array}
\right)
,~~~\label{bdgbdg}
\end{equation}
with
\begin{equation}
\hat{\xi}u_n(r_i)=-t\sum_{\hat{\delta}} 
u_n(r_i+\hat{\delta})+(\mu^I(r_i)+\mu)u_n(r_i),~~~\label{bdgxi}
\end{equation}

\begin{equation}
\hat{\Delta}u_n(r_i)=\Delta_0(r_i)u_n(r_i)+\sum_{\hat{\delta}} 
\Delta_{\delta}(r_i)u_n(r_i+\hat{\delta}),~~~\label{bdgdelta}
\end{equation}
where the gap functions are defined by

\begin{equation}
\Delta_0(r_i)\equiv 
V_0<c_{\uparrow}(r_i)c_{\downarrow}(r_i)>,~~~\label{bdgdelta0}
\end{equation}

\begin{equation}
\Delta_{\delta}(r_i)\equiv 
V_1<c_{\uparrow}(r_i+\hat{\delta})c_{\downarrow}(r_i)>,~~~\label{bdgdeltadelta}
\end{equation}
where $\hat{\delta}=\hat{x},-\hat{x},\hat{y},-\hat{y}$. Equations 
(\ref{bdgbdg}) are subject to the self consistency requirements 
\begin{equation}
\Delta_0(r_i)=V_0\sum_{n} 
u_n(r_i)v_n^{\ast}(r_i)\tanh\left(\frac{\beta 
\epsilon_n}{2}\right),~~~\label{bdgselfD0}
\end{equation}

\begin{eqnarray}
\Delta_{\delta}(r_i) & = & \frac{V_1}{2}\sum_{n} 
(u_n(r_i+\hat{\delta})v_n^{\ast}(r_i) \nonumber \\
 & + & u_n(r_i)v_n^{\ast}(r_i+\hat{\delta}) )\tanh\left(\frac{\beta 
\epsilon_n}{2}\right) .~~~\label{bdgselfDdelta}
\end{eqnarray}

We start with the approximate initial conditions for the gap functions
(\ref{bdgselfD0},\ref{bdgselfDdelta}). After exact diagonalization of Eq. 
(\ref{bdgbdg}) 
we obtain the $u(r_i)$ and
$v(r_i)$ and the eigenenergies $\epsilon_n$.
The quasiparticle amplitudes are then inserted into Eq. 
(\ref{bdgselfD0},\ref{bdgselfDdelta}) and new 
gap functions $\Delta_0(r_i)$ and $\Delta_{\delta}(r_i)$ are evaluated. 
We reinsert these quantities into Eq. (\ref{bdgxi},\ref{bdgdelta}),
and we proceed in the same way until we achieve selfconsistency,    
i.e., when the norm of the difference of $\Delta_0(r_i)$ and 
$\Delta_{\delta}(r_i)$ from their previous values  
is less than the desired accuracy.
We then compute the $d$-wave and the extended $s$-wave gap 
functions given by the expressions \cite{kallin1}:
\begin{equation}
\Delta_d(r_i)=\frac{1}{4}[\Delta_{\hat{x}}(r_i)+\Delta_{-\hat{x}}(r_i)
-\Delta_{\hat{y}}(r_i)-\Delta_{-\hat{y}}(r_i)],~~~\label{bdgdeltad}
\end{equation}

\begin{equation}
\Delta_s^{ext}(r_i)=\frac{1}{4}[\Delta_{\hat{x}}(r_i)+\Delta_{-\hat{x}}(r_i)
+\Delta_{\hat{y}}(r_i)+\Delta_{-\hat{y}}(r_i)].~~~\label{bdgdeltas}
\end{equation}
The number density at the $i$th site is given by
\begin{equation}
n_i=n_{i\uparrow}+n_{i\downarrow}=\sum_{n} 
\left [ |u_n(r_i)|^2 f(\epsilon_n) 
+ |v_n(r_i)|^2 (1-f(\epsilon_n)) \right ]
,~~~\label{bdgnumber}
\end{equation}
and the local density of states (LDOS) at the $i$th site is given by
\begin{equation}
\rho_i(E)=-2\sum_{n} 
\left [ |u_n(r_i)|^2 f^{'}(E-\epsilon_n) 
+ |v_n(r_i)|^2 f^{'}(E+\epsilon_n) \right ]
,~~~\label{bdgdos}
\end{equation}
where the factor $2$ comes from the twofold spin degeneracy,  
$f^{'}$ is the derivative of the Fermi function,
\begin{equation}
f(\epsilon)=\frac{1}{\exp(\epsilon/k_B T) + 1}
\end{equation}.

\section{Corner of superconductor}

\begin{figure}
\begin{center} 
\leavevmode 
\centerline{
\psfig{figure=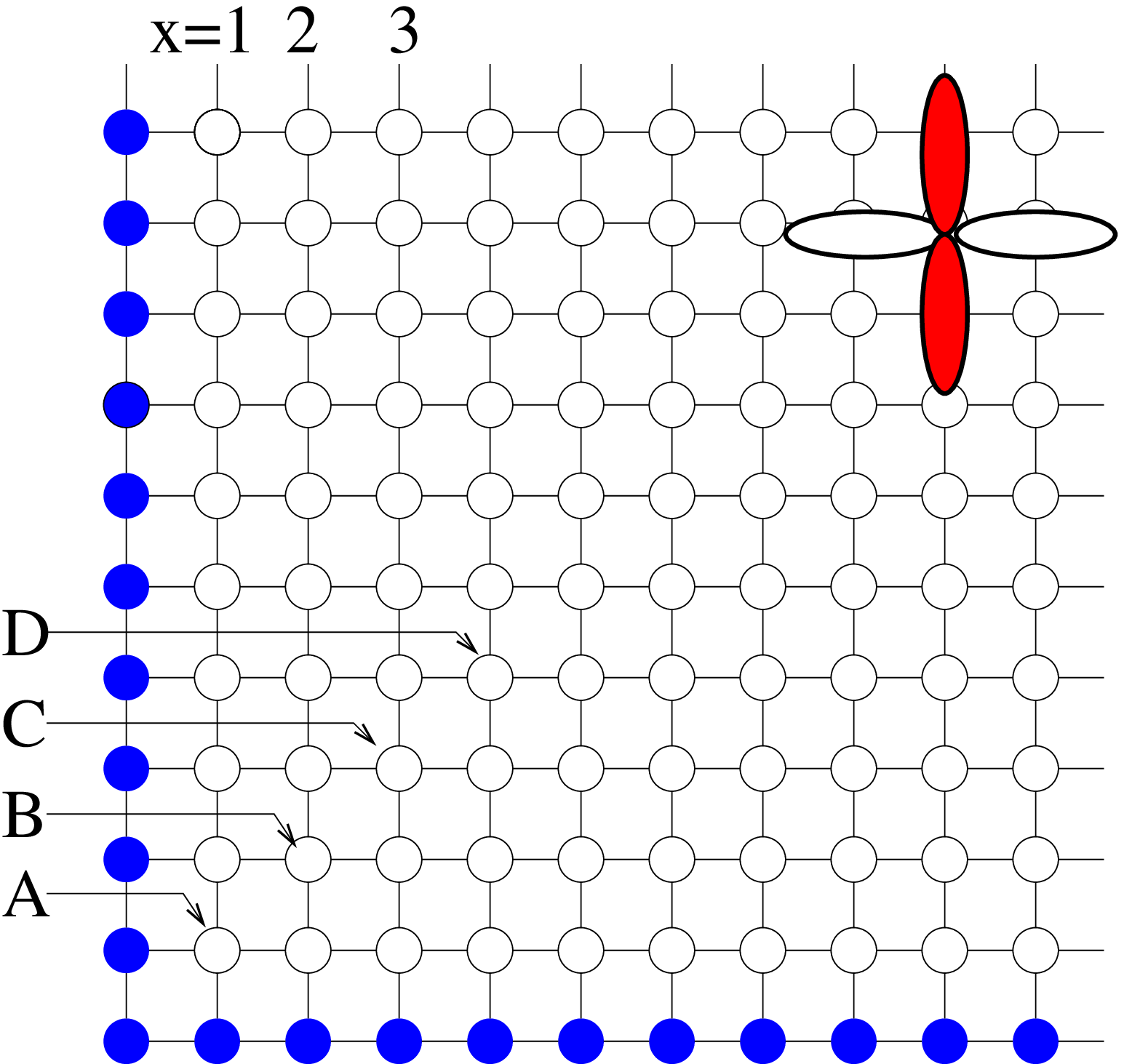,width=4.5cm,angle=0}}
\end{center}
\caption{The square lattice close to the corner. The chemical potential 
is set to $\mu^I=100t$ at the shaded surface sites.} 
\label{lattice}
\end{figure}

In this section we study 
the spatial variation of the order parameter close to the corner 
surface of a two dimensional square lattice seen in Fig. \ref{lattice}. 
The different symmetries 
are introduced by varying the strength of the local and 
non-local pairing interaction 
constants $V_0$, $V_1$.

\begin{figure}
\begin{center} 
\leavevmode 
\centerline{
\psfig{figure=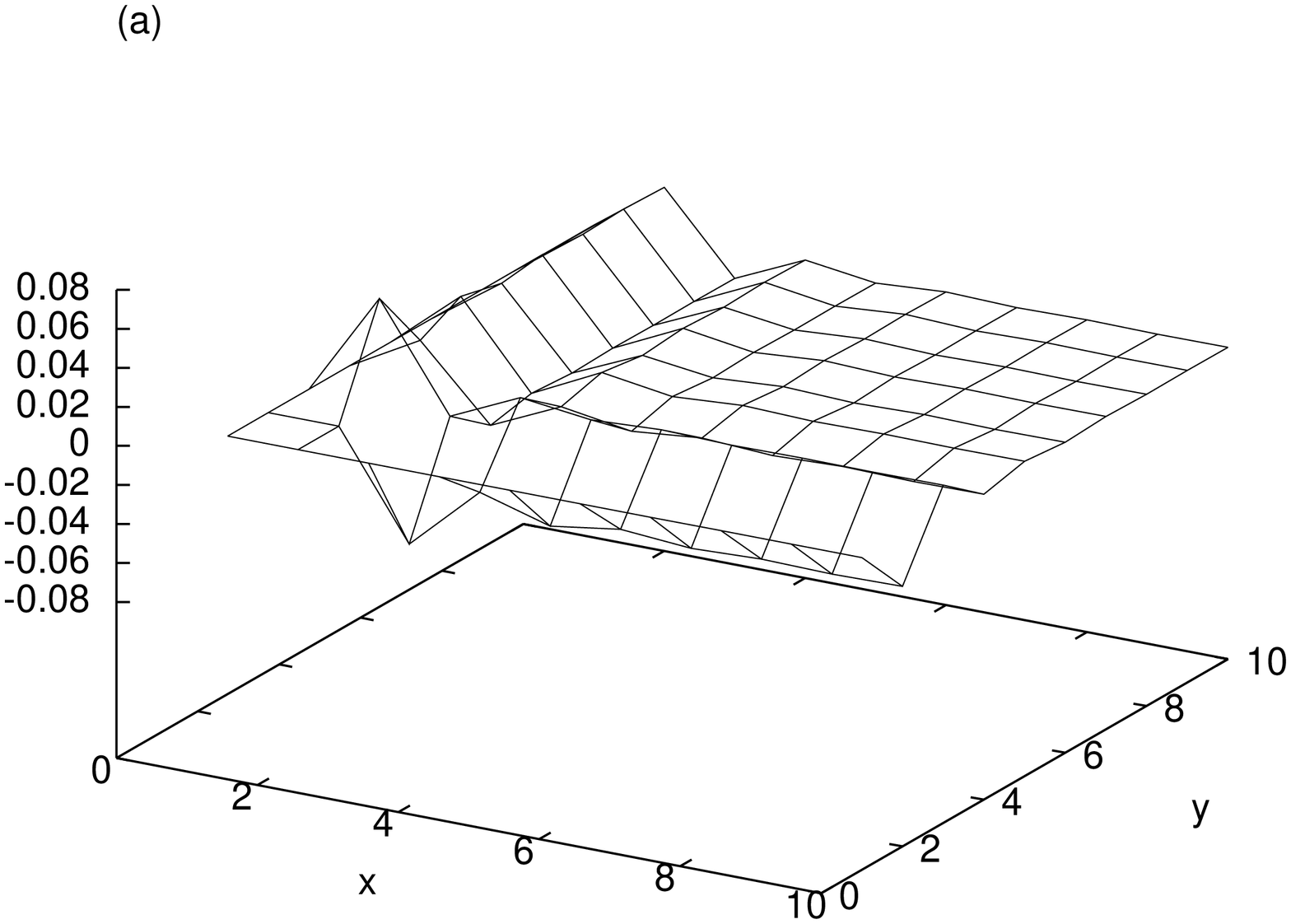,width=8.5cm,angle=-90}
            } 
\centerline{
\psfig{figure=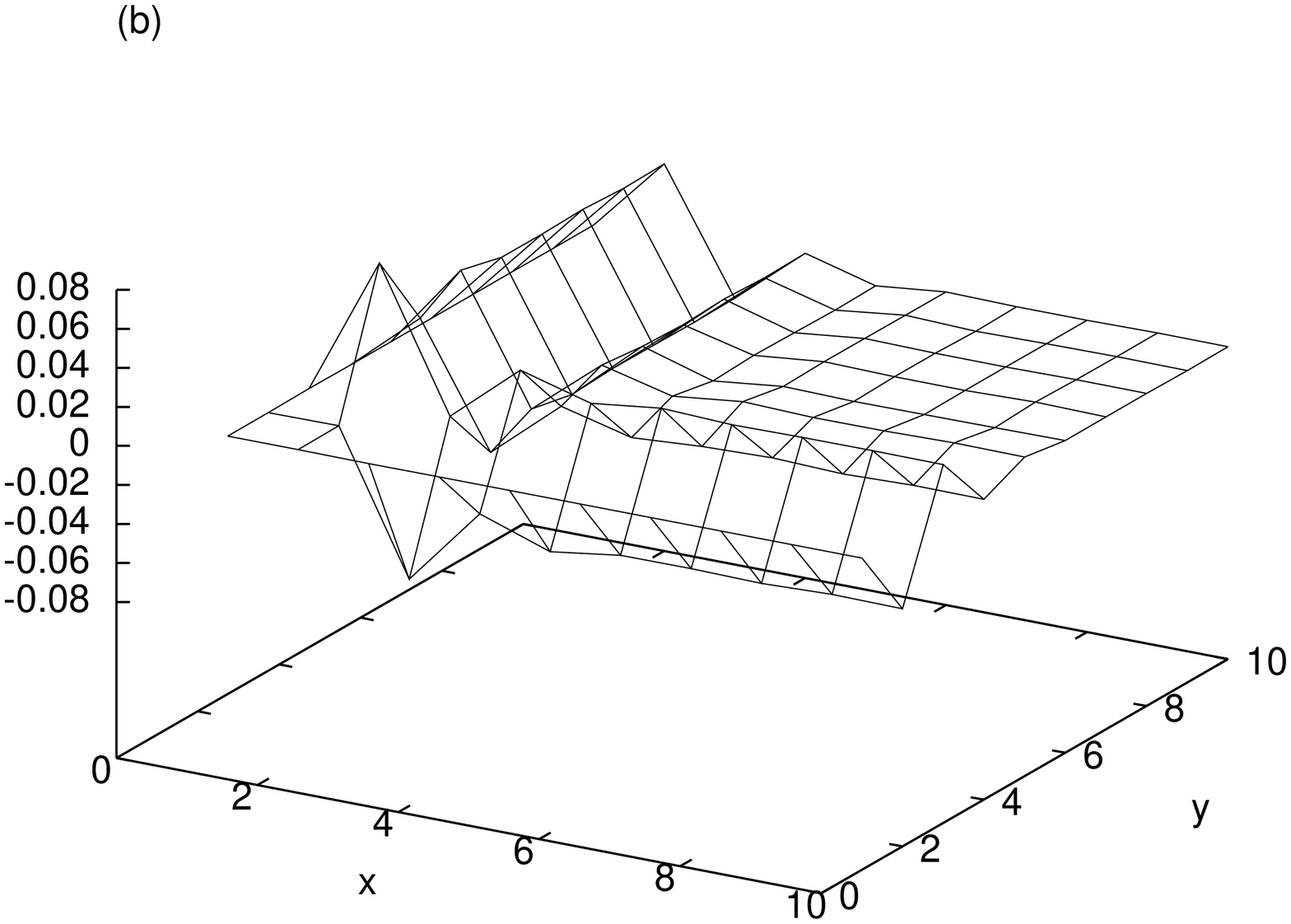,width=8.5cm,angle=-90}
}
\end{center}
\caption{Spatial dependence of the extended $s$-wave $\Delta_s^{ext}$
order parameter close to the corner of a two dimensional square lattice.
(a) $\mu=0$, (b) $\mu=t$.} 
\label{bulkopsext}
\end{figure}

\begin{figure}
\begin{center} 
\leavevmode 
\centerline{
\psfig{figure=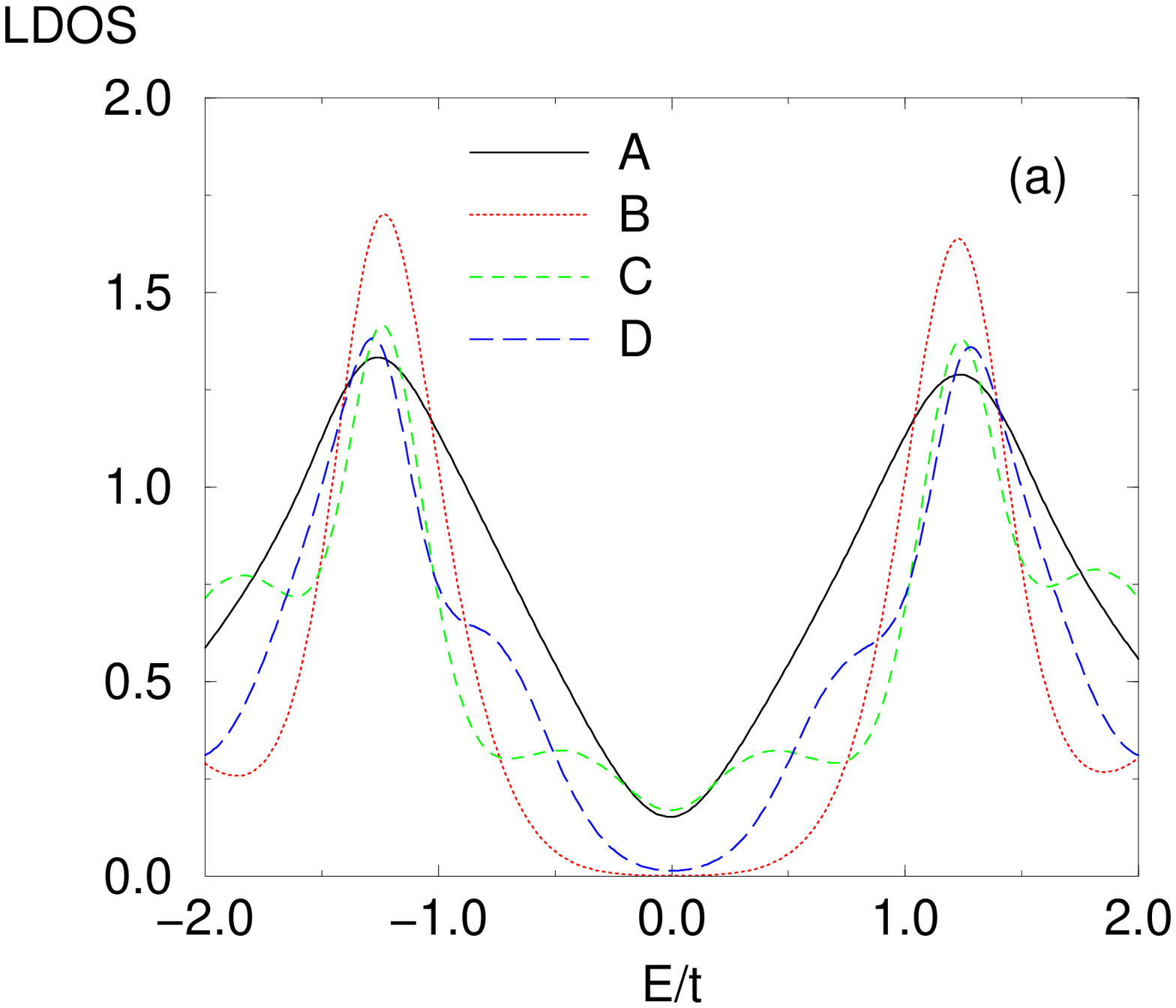,width=8.5cm}}
\centerline{
\psfig{figure=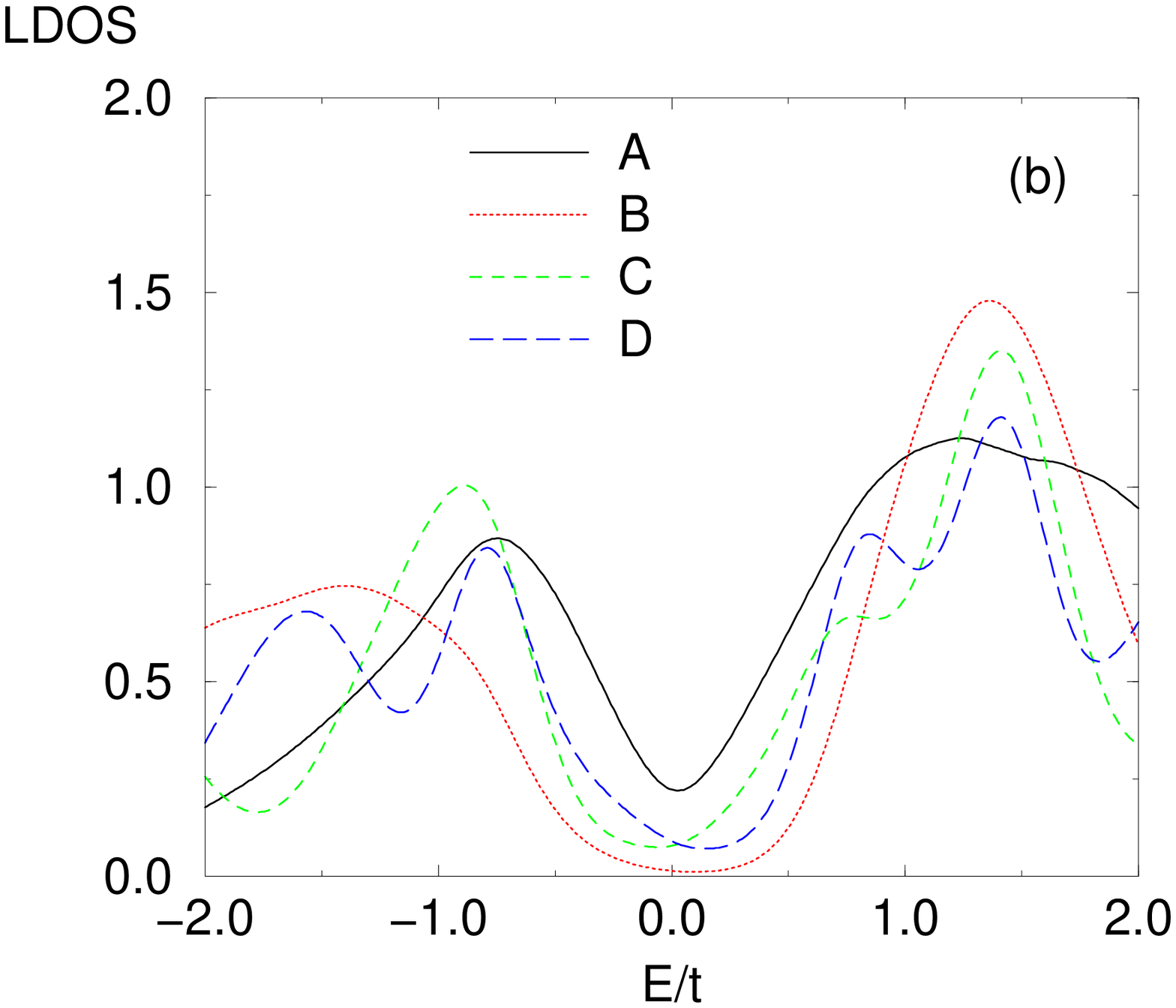,width=8.5cm}}
\end{center}
\caption{The local density of states at sites $A, B, C, D$ 
shown in Fig. 1, along the 
diagonal of the two dimensional square lattice. (a) $\mu=0$, (b) $\mu=t$. 
}  
\label{bulkdos}
\end{figure}

\begin{figure}
\begin{center} 
\leavevmode 
\centerline{
\psfig{figure=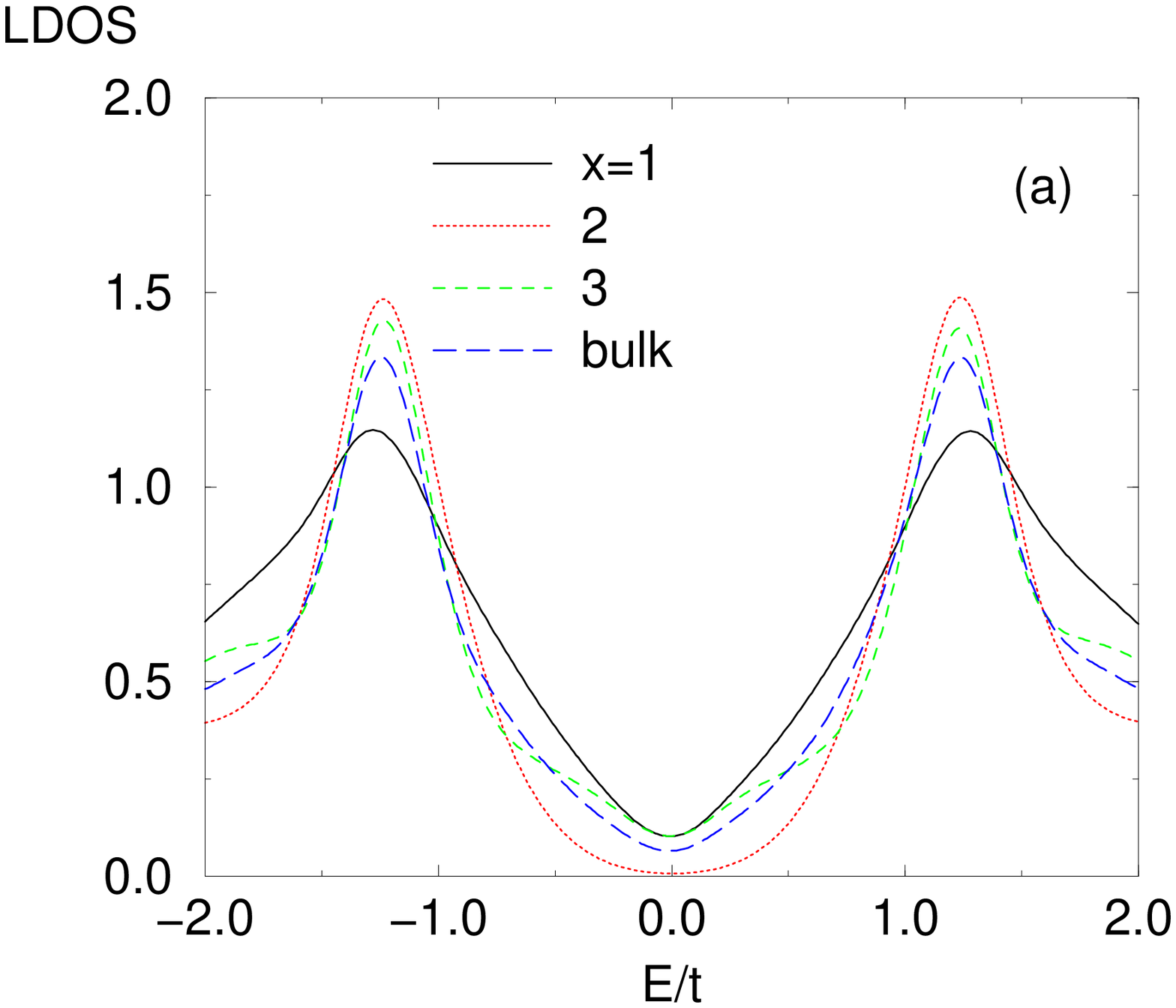,width=8.5cm}}
\centerline{
\psfig{figure=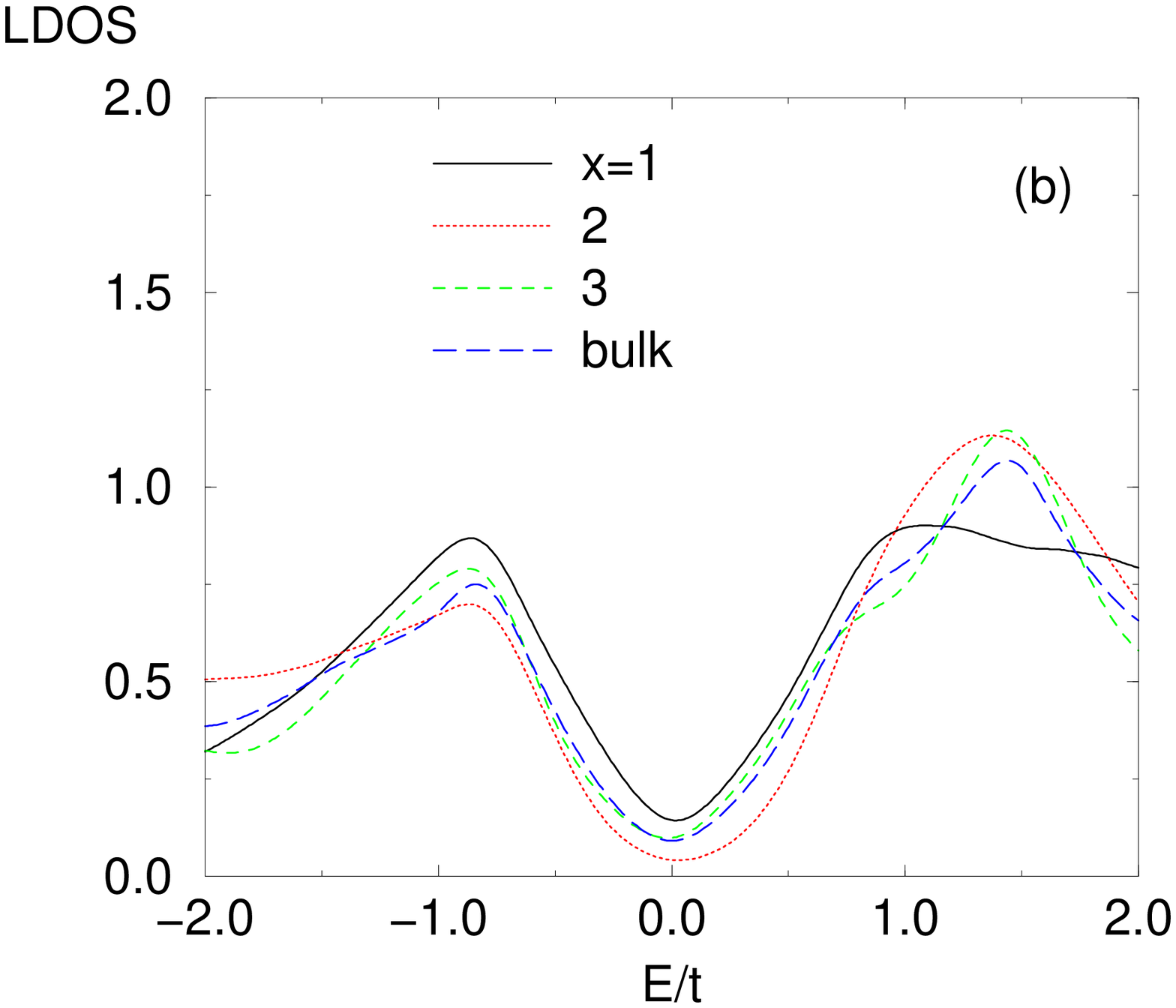,width=8.5cm}}
\end{center}
\caption{
The local density of states at sites $x=1,2,3$ shown in Fig. 1
along the $[100]$ surface of the two dimensional square lattice, 
and the bulk density of states. (a) $\mu=0$, (b) $\mu=t$.
}  
\label{dos100}
\end{figure}

We consider a two dimensional system of $30\times 30$ sites, and we suppose 
fixed boundary conditions by setting the impurity potential $\mu^{I}=100t$
at the surface.
The temperature is $k_B T=0.1t$.
The constants  $V_0$ and 
$V_1$ are $0.0$ and $-2.5t$ respectively. This choice of parameters 
gives $d$-wave superconductivity.
The $\Delta_d$ order parameter is shown 
in Fig. ~\ref{bulkopd}, 
for two different values of the chemical potential $\mu=0$ 
and $\mu=t$.
For $\mu=0$ the bulk value of the gap is $\Delta_d=0.32t$, while 
the transition temperature is calculated as $k_BT_c=0.6t$.
It is much more suppressed for finite values of the chemical potential.
As we can see the $d$-wave order parameter is enhanced near the surface 
from its bulk value, 
and goes to zero at the surface atoms because the hopping to these sites 
is difficult due to the impurity barrier.
In Fig. \ref{bulkopsext} we see the induced 
extended $s$-wave order parameter $\Delta_s^{ext}$, 
for two different values of the chemical potential $\mu=0$ and $\mu=t$. 
It is much more enhanced for greater values of the chemical potential. 
Near the surface $\Delta_s^{ext}$ oscillates 
at an atomic scale and vanishes into the 
bulk region at a distance of few lattice sites. 
It reverses its sign on either side of the lattice 
edge and it is exactly zero in the diagonal direction. 
Next to the corner we see an enhancement from the edge value.
It appears to have a $d$-wave 
like structure just at the corner of the square lattice. This behavior 
is also seen near impurities \cite{kallin2}, and across twin boundaries 
\cite{kallin3} using BdG equations within the extended Hubbard model in a two 
dimensional orthorhombic lattice. 
The explanation beyond this is the sign change of the 
$d$-wave order parameter across the $[110]$ direction close to the 
corner. The quasiparticles feel the sign change of the pairing
potential when they are reflected from the lattice edges near the 
corner.
In Fig. ~\ref{bulkdos}(a) we plot the local density of states 
for $\mu=0$ at points 
$A, B, C, D$
seen in Fig. ~\ref{lattice}, along the diagonal of the lattice. 
The symmetry with respect to $E=0$ is due to the zero chemical 
potential.
The LDOS is site depended and shows a complicated gap structure.
Also no ZBCP has been observed in 
agreement with the results of the quasiclassical theory \cite{stefan1}.
As we move to the interior of 
the lattice the LDOS converges to the bulk density of states in 
a two dimensional square lattice.
When $\mu$ deviates from zero as in Fig. ~\ref{bulkdos}(b) the 
LDOS becomes asymmetric. However no ZBCP is formed.
In Fig. ~\ref{dos100}(a) we plot the local density of states
for $\mu=0$,  at sites $x=1,2,3$ 
seen in Fig. ~\ref{lattice}, along the direction $[100]$ perpendicular 
to the lattice edge. 
The LDOS has the $V$-like line shape, is symmetric with respect 
to $E=0$ due to the electron-hole symmetry, and has the minimum at $E=0$ 
which characterizes the $d$-wave 
symmetry of the order parameter. 
It is seen that the LDOS recovers its 
bulk value (long dashed line) as we move away from the surface.
The symmetric form of the LDOS line shape is lost when the 
chemical potential deviates from zero, as seen in Fig. \ref{dos100}(b) 
for $\mu=t$. 

\begin{figure}
\begin{center}
\leavevmode
\centerline{
\psfig{figure=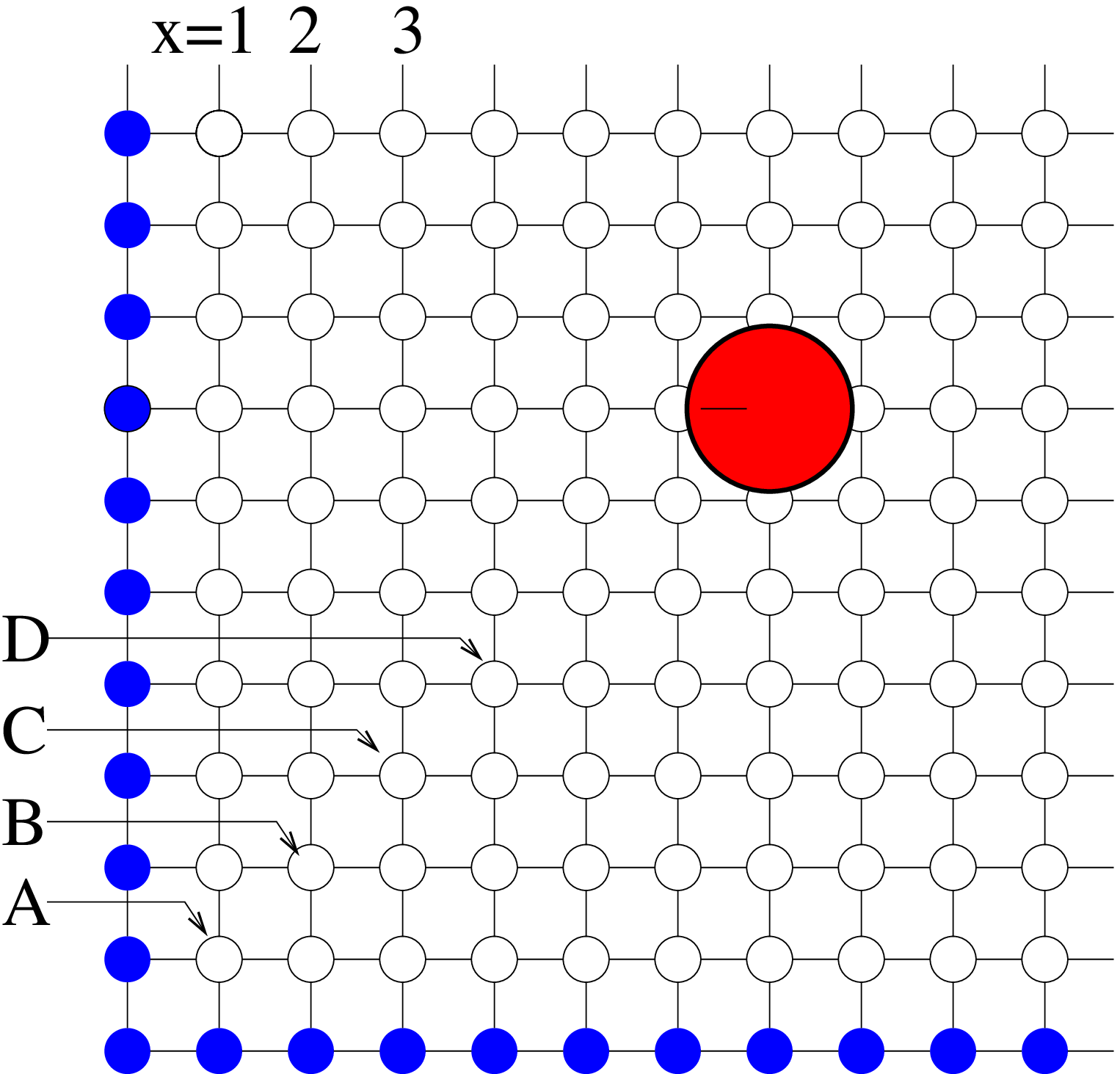,width=4.5cm,angle=0}}
\end{center}
\caption{The square lattice close to the corner. The chemical potential
is set to $\mu^I=100t$ at the shaded surface sites.}
\label{lattices}
\end{figure}

\begin{figure}
\begin{center} 
\leavevmode 
\centerline{
\psfig{figure=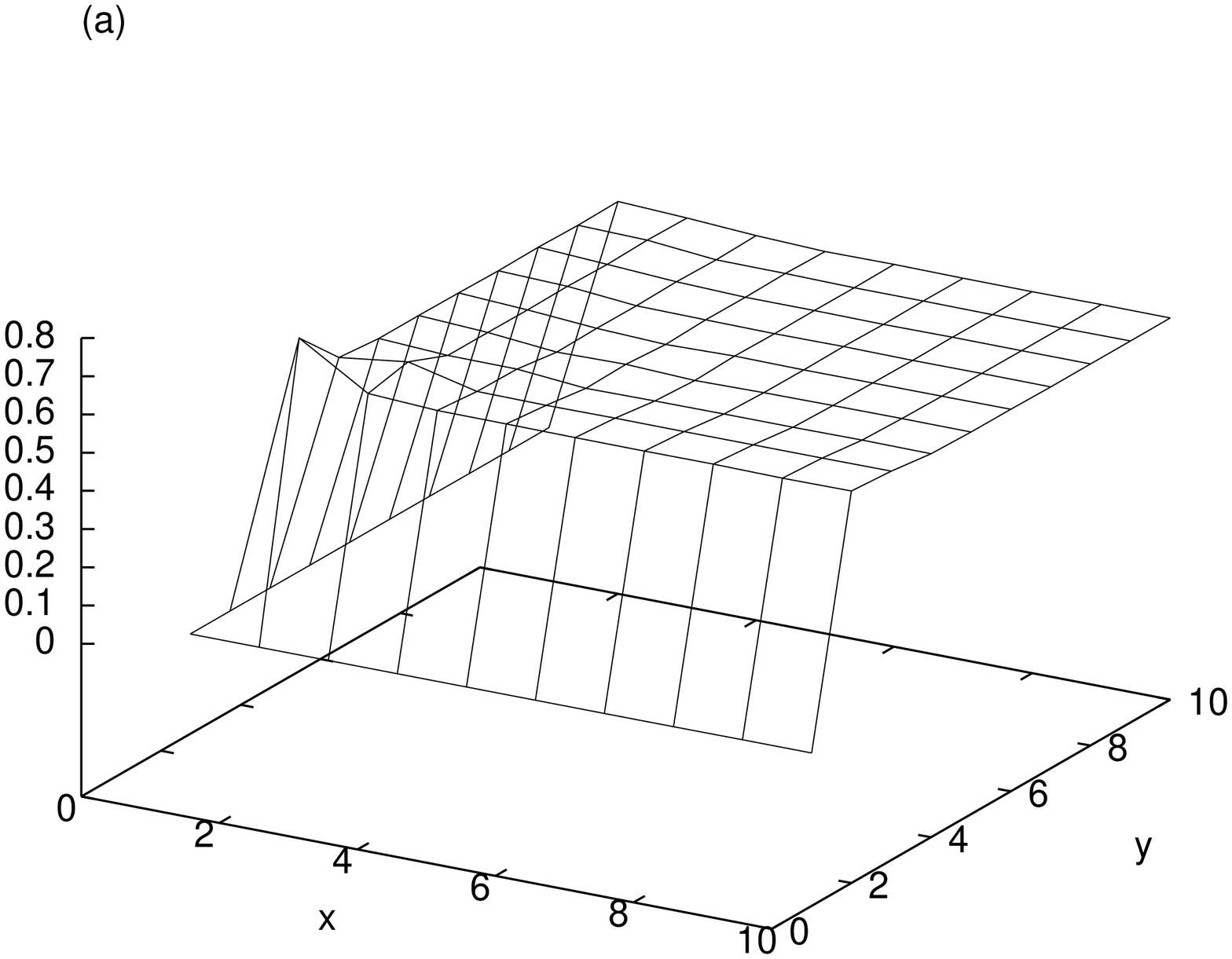,width=8.5cm,angle=-90}
}
\centerline{
\psfig{figure=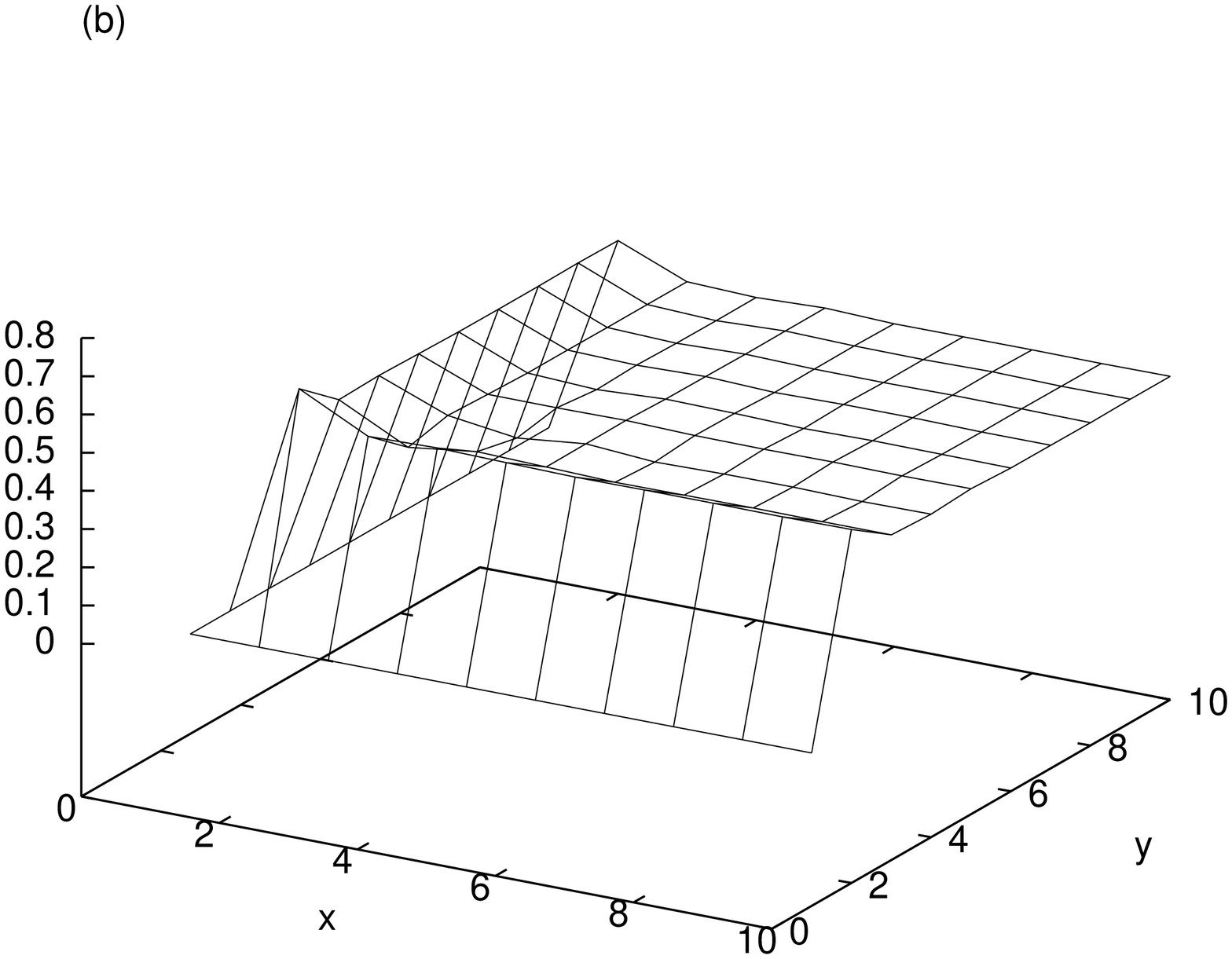,width=8.5cm,angle=-90}
}
\end{center}
\caption{Spatial dependence of the $s$-wave  
order parameter close to the corner of a two dimensional square lattice.
(a) $\mu=0$, (b) $\mu=t$.} 
\label{bulkops}
\end{figure}

\begin{figure}
\begin{center} 
\leavevmode 
\centerline{
\psfig{figure=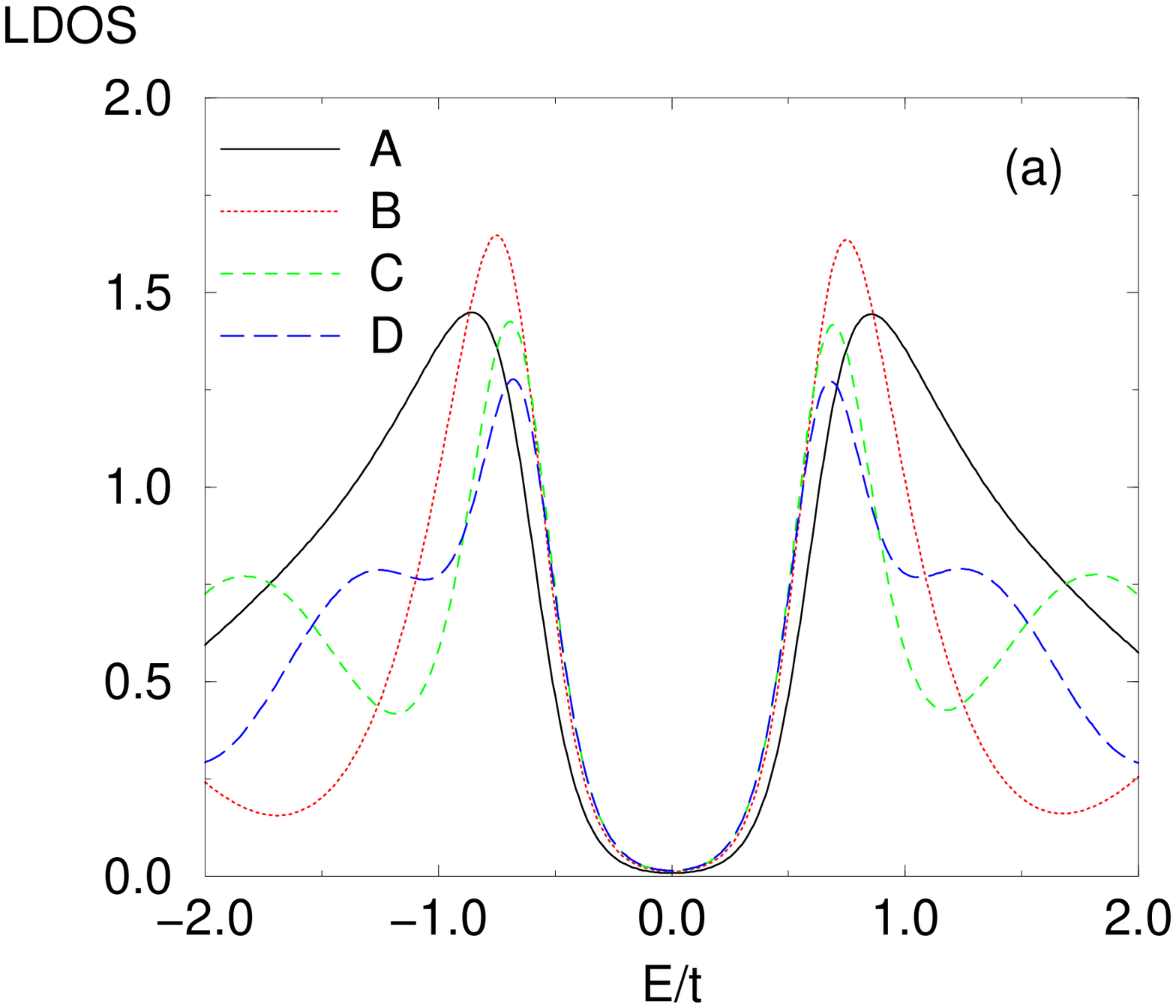,width=8.5cm}}
\centerline{
\psfig{figure=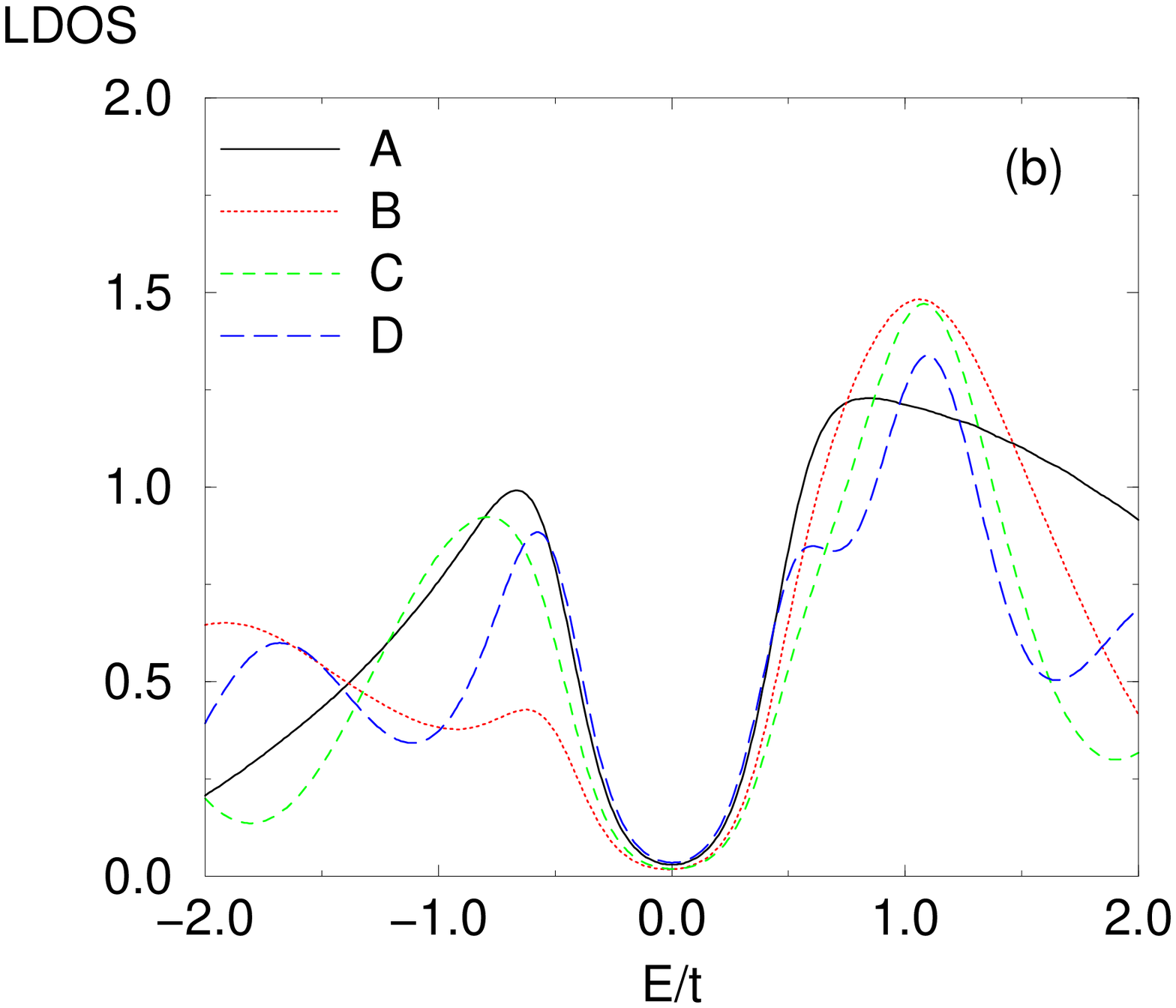,width=8.5cm}}
\end{center}
\caption{The local density of states at sites $A, B, C, D$ 
shown in Fig. 6 along the 
diagonal of the two dimensional square lattice.  
The pairing symmetry is $s$. (a) $\mu=0$, (b) $\mu=t$.
}  
\label{bulkdoss}
\end{figure}

\begin{figure}
\begin{center} 
\leavevmode 
\centerline{
\psfig{figure=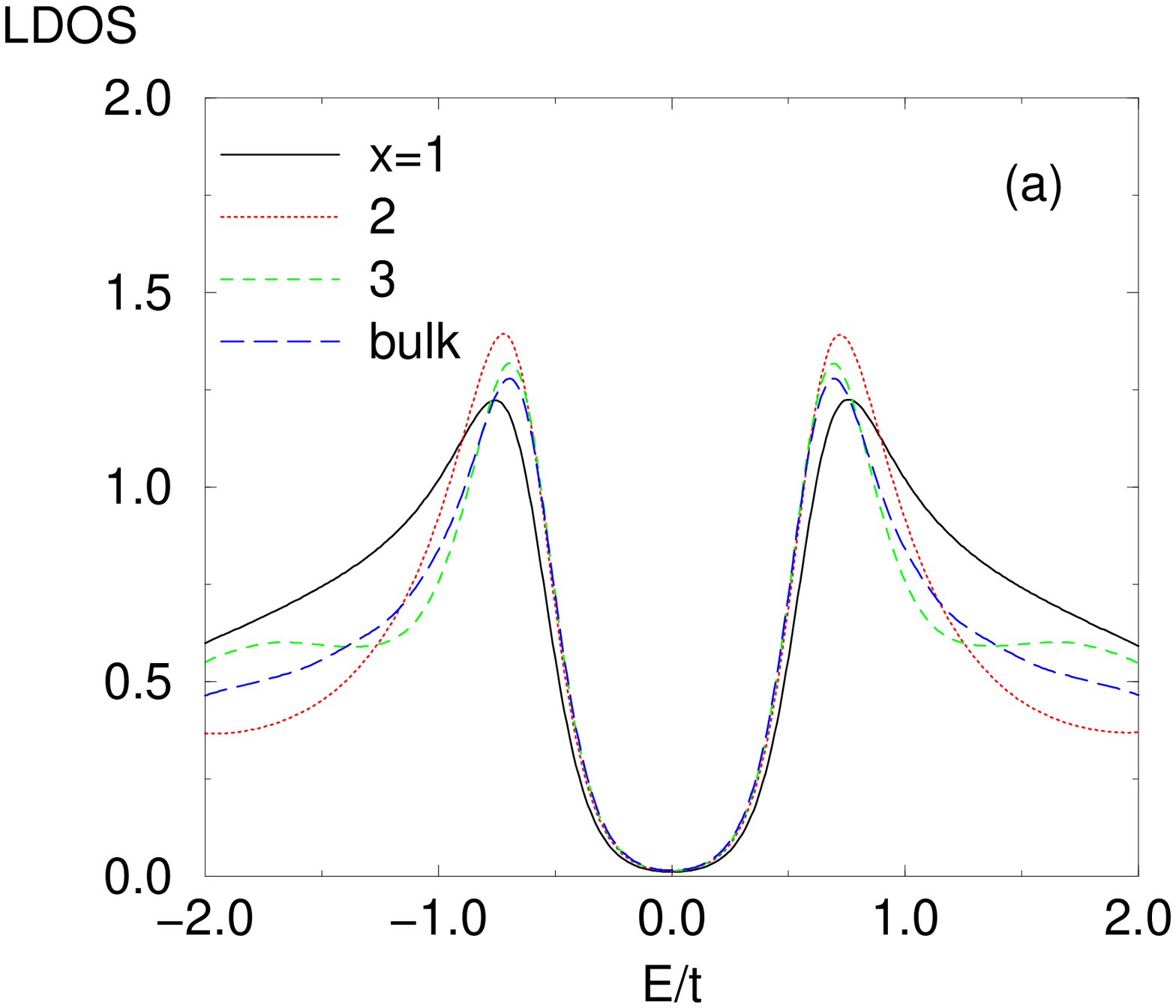,width=8.5cm}}
\centerline{
\psfig{figure=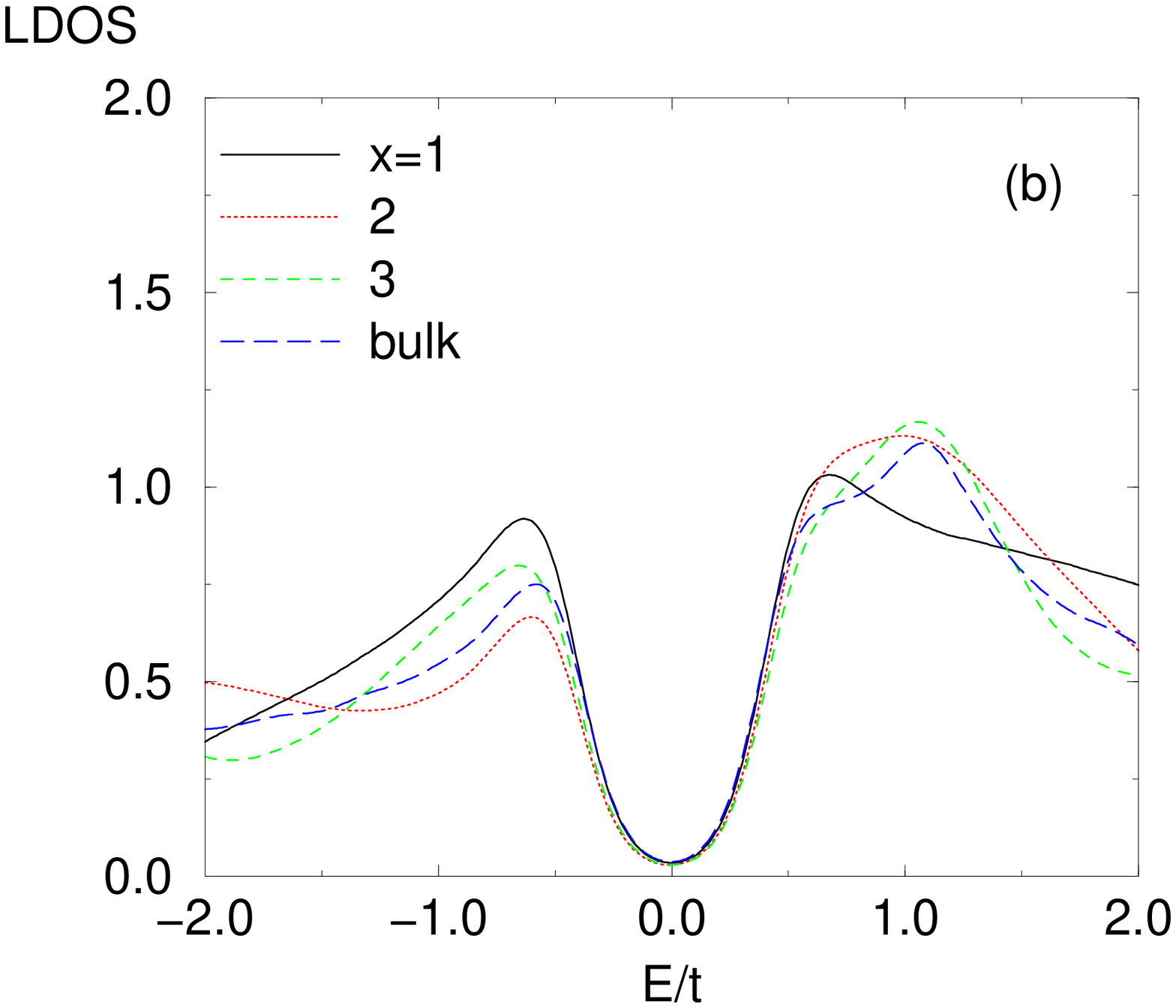,width=8.5cm}}
\end{center}
\caption{
The local density of states at sites $x=1,2,3$ 
shown in Fig. 6 along 
the $[100]$ surface of the two dimensional square lattice, 
and the bulk density of states. The pairing symmetry is $s$.
(a) $\mu=0$, (b) $\mu=t$.
}  
\label{dos100s}
\end{figure}

To understand the effect of the different symmetry  
the local pairing interaction 
is set to the value $V_0=-2.5t$, while the non-local interaction 
is set to the value $V_1=0$. This set of parameters gives rise to 
$s$-wave pairing. The spatial variation of the $s$-wave order parameter
close to the corner of a two dimensional square lattice 
(see Fig. \ref{lattices}) is seen in Fig. \ref{bulkops}, 
for two different values of the chemical potential $\mu=0$ 
and $\mu=t$. 
For $\mu=0$ as seen in Fig. \ref{bulkops}(a) 
the order parameter increases monotonically 
(which is to be compared to the $d$-wave case where this increase is 
non monotonous) to the 
bulk value at the sides of the lattice while at the corner an enhancement,
relative to the bulk value  
is seen which disappears over a few sites to the interior of the lattice. 
For $\mu=t$ the order parameter increases non-monotonically even 
at the sides of the lattice as seen in Fig. \ref{bulkops}(b).
In Fig. \ref{bulkdoss}a (b) 
we plot the local density of states at points $A, B, C, D$, 
for $\mu=0$($\mu=t$) 
seen in Fig. ~\ref{lattices}, along the diagonal direction 
of the lattice.
The LDOS shows gap structure with $U$-like line shape 
which characterizes the $s$-wave symmetry and is symmetric 
(asymmetric) when $\mu$ is (deviates from) zero . The LDOS is insensitive 
to the orientation of the surface and is site independent.
Similar behavior is seen for the $[100]$ surface seen in 
Fig. \ref{dos100s}.

The absence of ZBCP in the LDOS can be 
explained from the results 
of the quasiclassical theory as follows. The condition for the 
formation of ZBCP is the change of sign of the quasiparticles during 
the scattering from the surface of the superconductor. In $s$-wave 
superconductors this sign change does not occur at surfaces or 
interfaces due to the isotropy of the pair potential. In anisotropic 
superconductors this sign change is possible for certain orientation 
of the surface. However for the corner surface, at the direction 
where the lobes of the $d$-wave order parameter are at right angles 
to the surface, a typical trajectory of a quasiparticle would consist
of two subsequent reflections from the lattices edges, in non of which 
sign change of the order parameter occurs. Therefore the quasiparticle 
does not feel the sign change of the order parameter and no ZBCP is 
formed. For a corner where the lobes are not exactly at right angles 
to the surface the condition for the formation of Andreev bound states 
at the surface can occur and also the ZBCP.   

\section{$s-I-n$, $d-I-n$ interfaces along the $[110]$ direction}

We now discuss the effect of the interface on the order parameter 
and the local density of states for different symmetries.
The interface is modeled by a line of sites along the diagonal of the 
lattice, $y'$ direction, where the chemical potential is set to a value 
in accordance with the strength of the barrier we want to model. 
The value of the interaction strength in each part of the interface 
determines the particular system that we are considering.

\begin{figure}
\begin{center} 
\leavevmode 
\centerline{
\psfig{figure=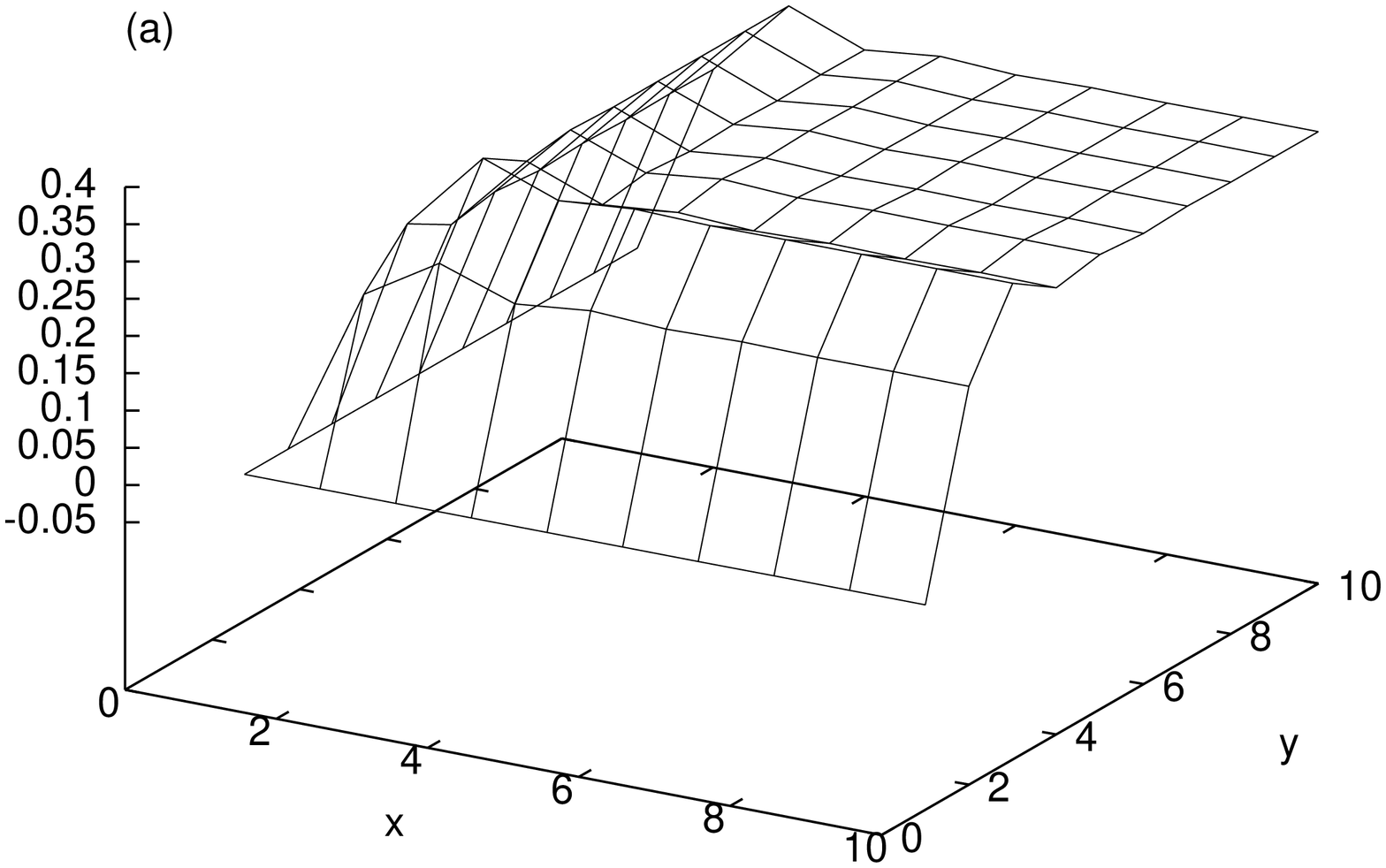,width=8.5cm,angle=-90}
            } 
\centerline{
\psfig{figure=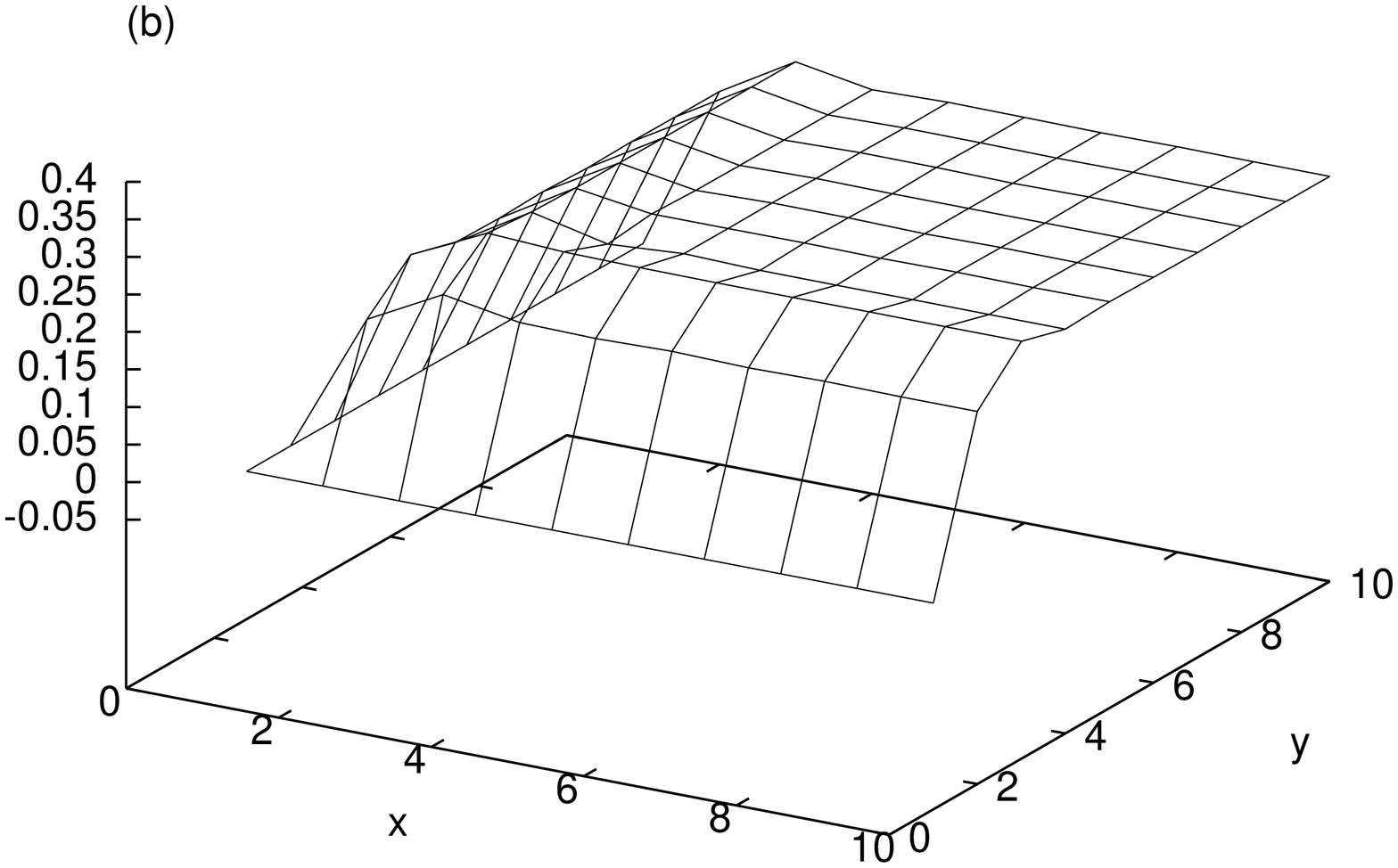,width=8.5cm,angle=-90}
}
\end{center}
\caption{Spatial dependence of the $d$-wave 
order parameter close to the corner of a two dimensional square lattice.
(a) $\mu=0$, (b) $\mu=t$.} 
\label{bulkopd}
\end{figure}

\begin{figure}
\begin{center} 
\leavevmode 
\centerline{
\psfig{figure=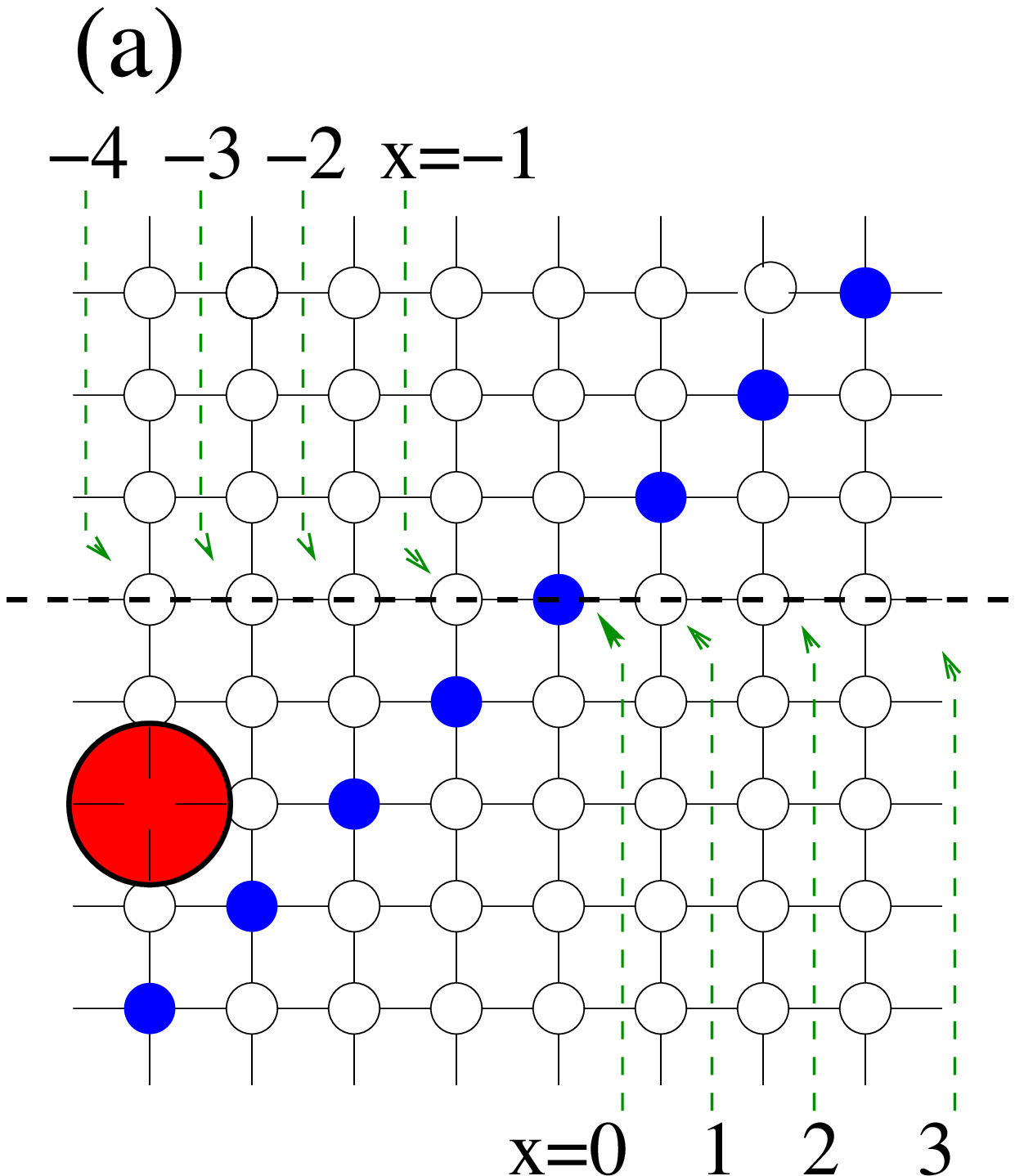,width=5.5cm,angle=0}}
\centerline{\hbox{
\psfig{figure=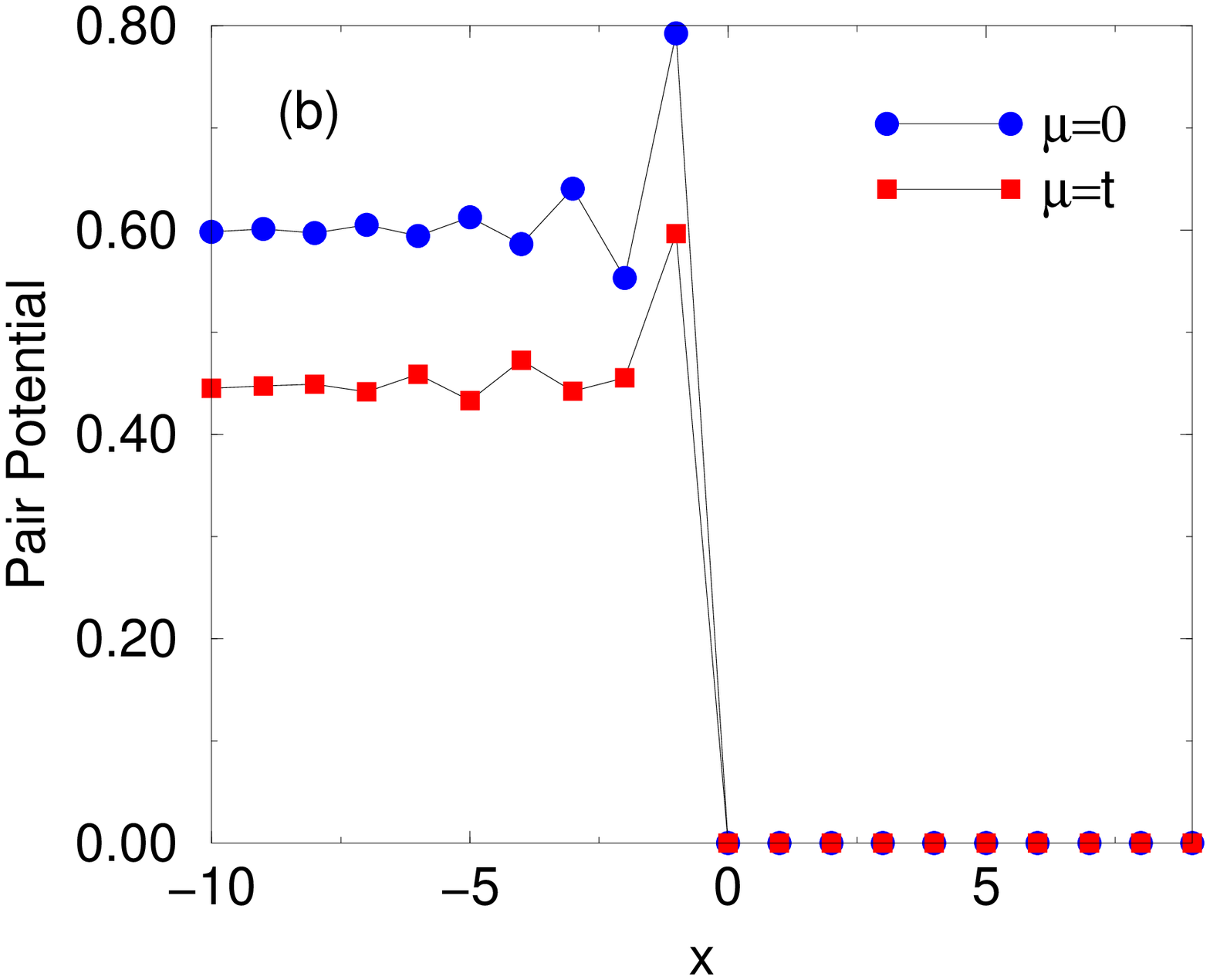,width=8.5cm}}}
\centerline{\hbox{
\psfig{figure=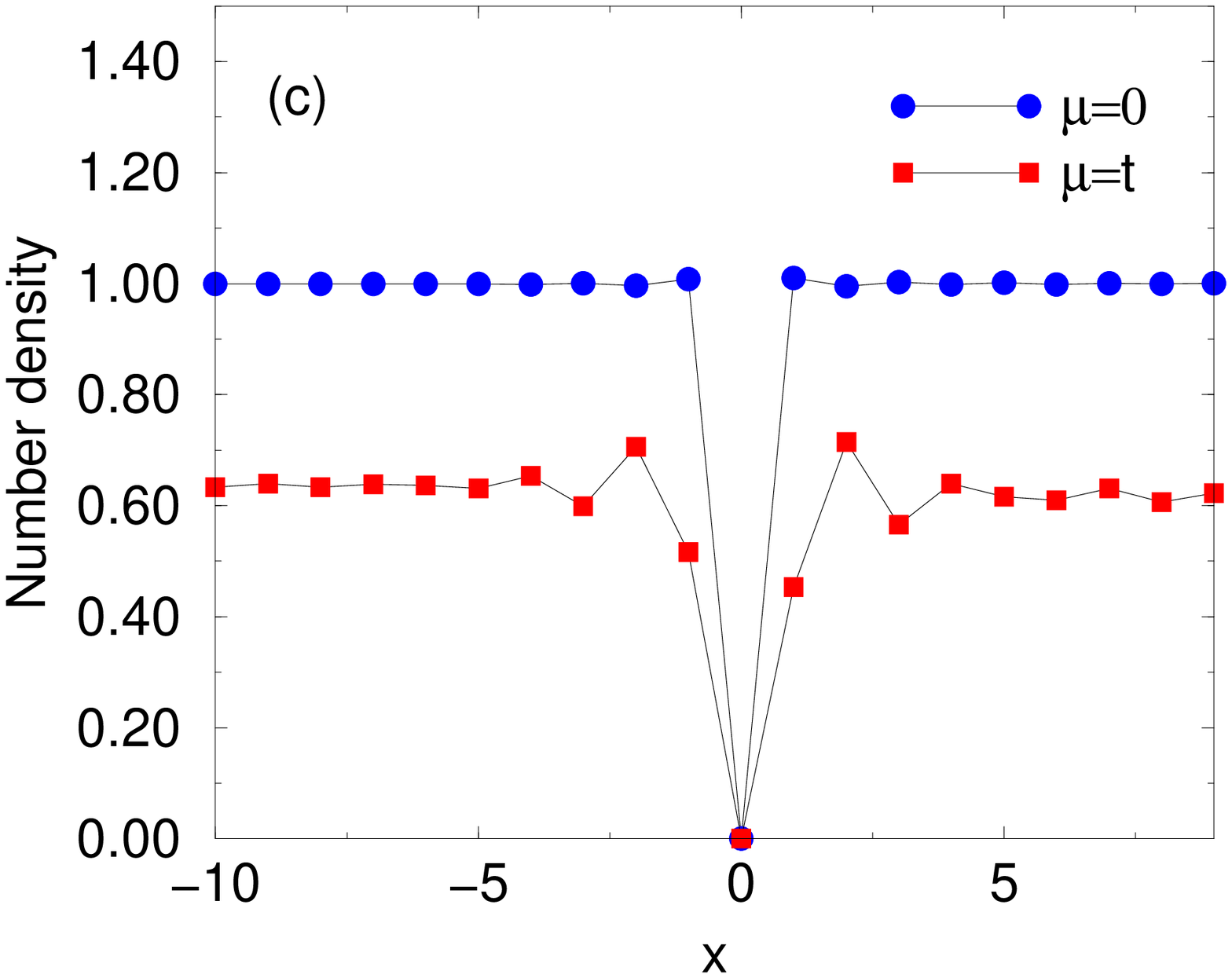,width=8.5cm}}}
\end{center}
\caption{
(a) Spatial distribution of impurities, indicated as solid
circles, corresponding to
a $s-I-n$ interface along the $[110]$ surface.
Also the labeling of the sites along the $x$ direction is shown.
(b) The magnitude of the $s$-wave component $\Delta_s$ of the 
superconducting order parameter as a function of $x$, for a $s-I-n$ 
interface for $\mu=0,t$. 
The order parameter is calculated  
along the thick dashed line in direction $x$ shown in (a).
(c) The number density $n_i$
as a function of $x$ shown in (a), for a $s-I-n$
interface for $\mu=0,t$.
}  
\label{sinop}
\end{figure}

To understand the effect of the symmetry of the pair potential 
we consider first the conventional $s$-wave superconductor-
insulator-normal metal ($s-I-n$) interface shown in Fig. \ref{sinop}(a). 
The local interaction 
in region where $x'<0$, ($x'$ is the direction perpendicular to the 
interface) is $V_0=-2.5t$, and the strength of the barrier 
is $\mu^{110}=100t$.
For $\mu=0$ the superconducting gap is $\Delta_s=0.6t$.
We plot in Fig. \ref{sinop}(b) the magnitude of the $s$-wave component 
$\Delta_s$ of the 
superconducting order parameter as a function of $x$ 
shown in Fig. \ref{sinop}(a) for two different values of the 
chemical potential, i.e., $\mu=0$ and $\mu=t$. 
It is suppressed near the interface and increases nonmonotonically 
to the bulk value at a few lattice sites. The enhancement at the topmost
sites close to 
the interface is similar to the spatial variation of the $d$-wave 
superconductor close to the $[100]$ surface seen in Fig. \ref{bulkopd}.
The bulk order parameter is suppressed when $\mu$ deviates from 0 and 
also the spatial oscillations close to the interface are of 
reduced amplitude as seen in Fig. \ref{sinop}(b). 
We also plot in Fig. \ref{sinop}(c) the number density $n_i$ for the electrons 
for the two different values of the chemical potential, 
i.e., $\mu=0$ and $\mu=t$. For $\mu=0$ the number density is unity 
(one electron per site) in the bulk and decays to zero at the interface.
However for finite $\mu$ the number density is reduced.

\begin{figure}
\begin{center} 
\leavevmode 
\centerline{
\psfig{figure=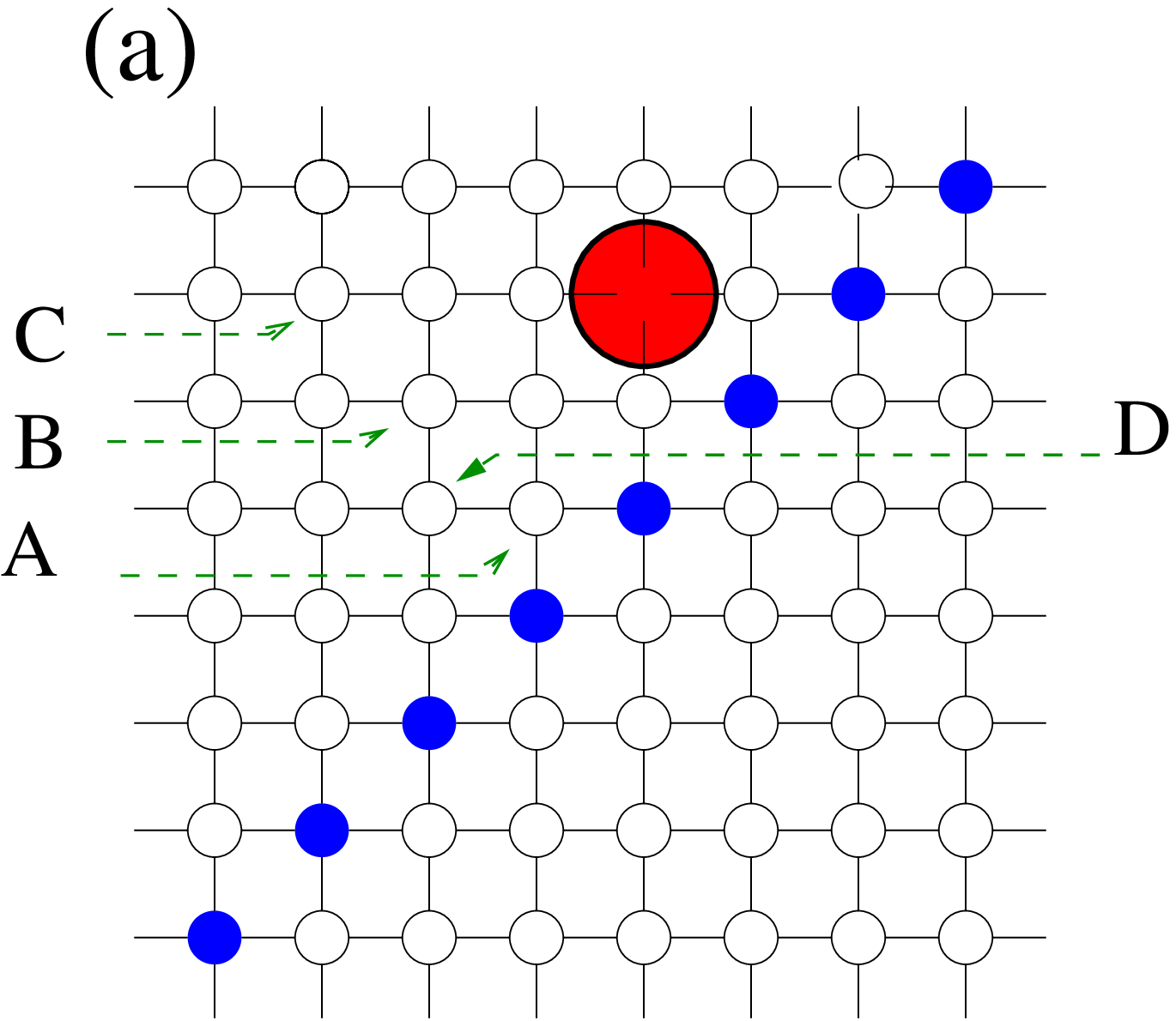,width=5.5cm,angle=0}}
\centerline{\hbox{
\psfig{figure=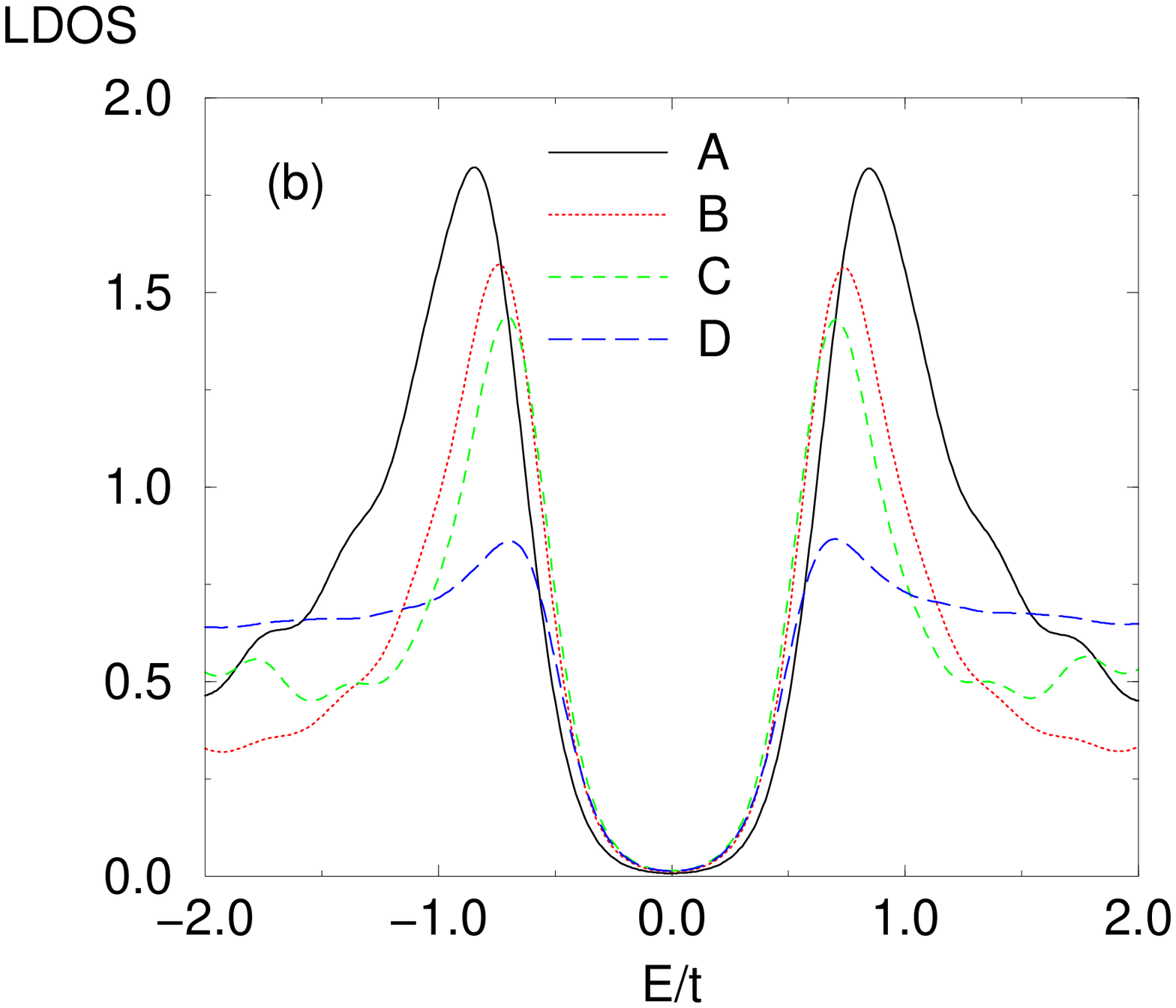,width=8.5cm}}}
\centerline{\hbox{
\psfig{figure=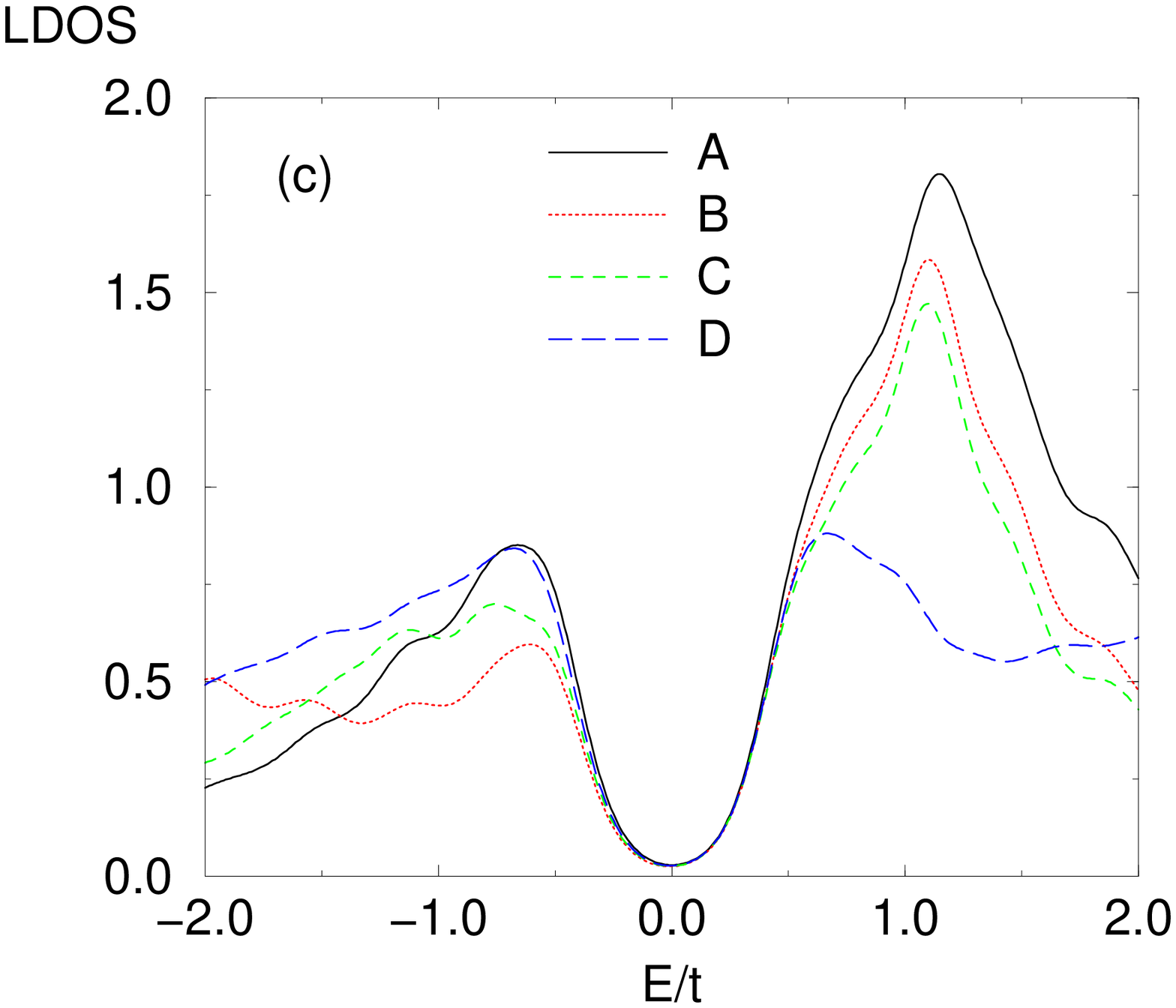,width=8.5cm}}}
\end{center}
\caption{
(a) Spatial distribution of impurities, indicated as solid 
circles, corresponding to 
a $s-I-n$ interface along the $[110]$ surface. 
(b) The local density of states (LDOS) at $A,B,C,D$, shown in (a) 
of a $s-I-n$ interface for $\mu=0$.  
(c) The local density of states (LDOS) at $A,B,C,D$, shown in (a) 
of a $s-I-n$ interface for $\mu=t$.  
}  
\label{sindos}
\end{figure}

In Fig. \ref{sindos}(b) we plot the local density of states (LDOS) at sites 
$A, B, C, D$, shown in Fig. \ref{sindos}(a), for $\mu=0$.
As expected the LDOS line shape is symmetric since $\mu=0$. 
The LDOS shows the gap structure and $U$-like line shape.
Also the LDOS is modified by the presence of the barrier, from it's bulk form.
The LDOS is insensitive to the direction of the interface.
For finite $\mu=1$ seen in Fig. \ref{sindos}(c) the LDOS keeps its 
$U$-like line shape. However it becomes asymmetric due to the 
breakdown of the electron-hole symmetry.  

\begin{figure}
\begin{center} 
\leavevmode 
\centerline{
\psfig{figure=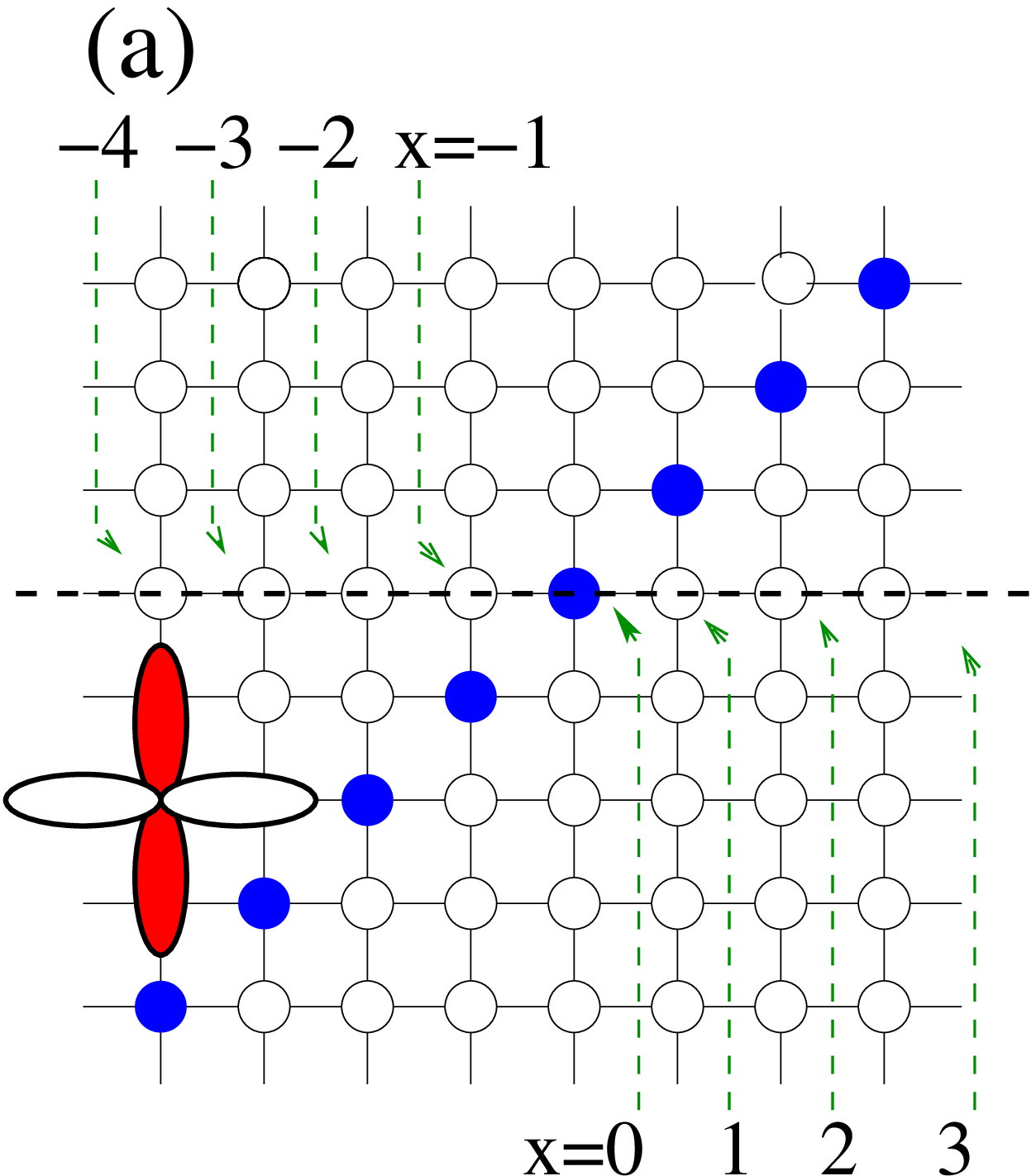,width=5.5cm,angle=0}}
\centerline{\hbox{
\psfig{figure=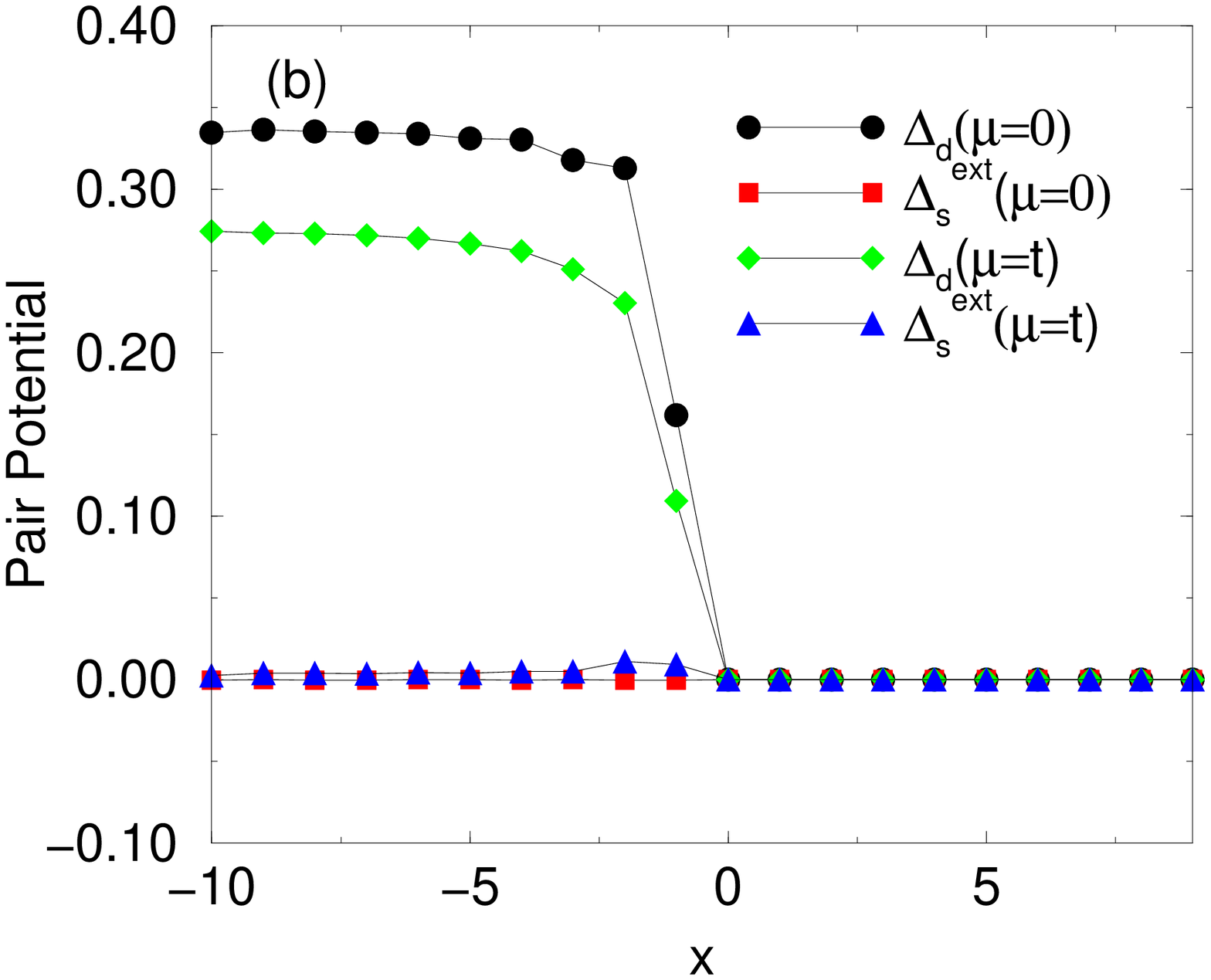,width=8.5cm}}}
\centerline{\hbox{
\psfig{figure=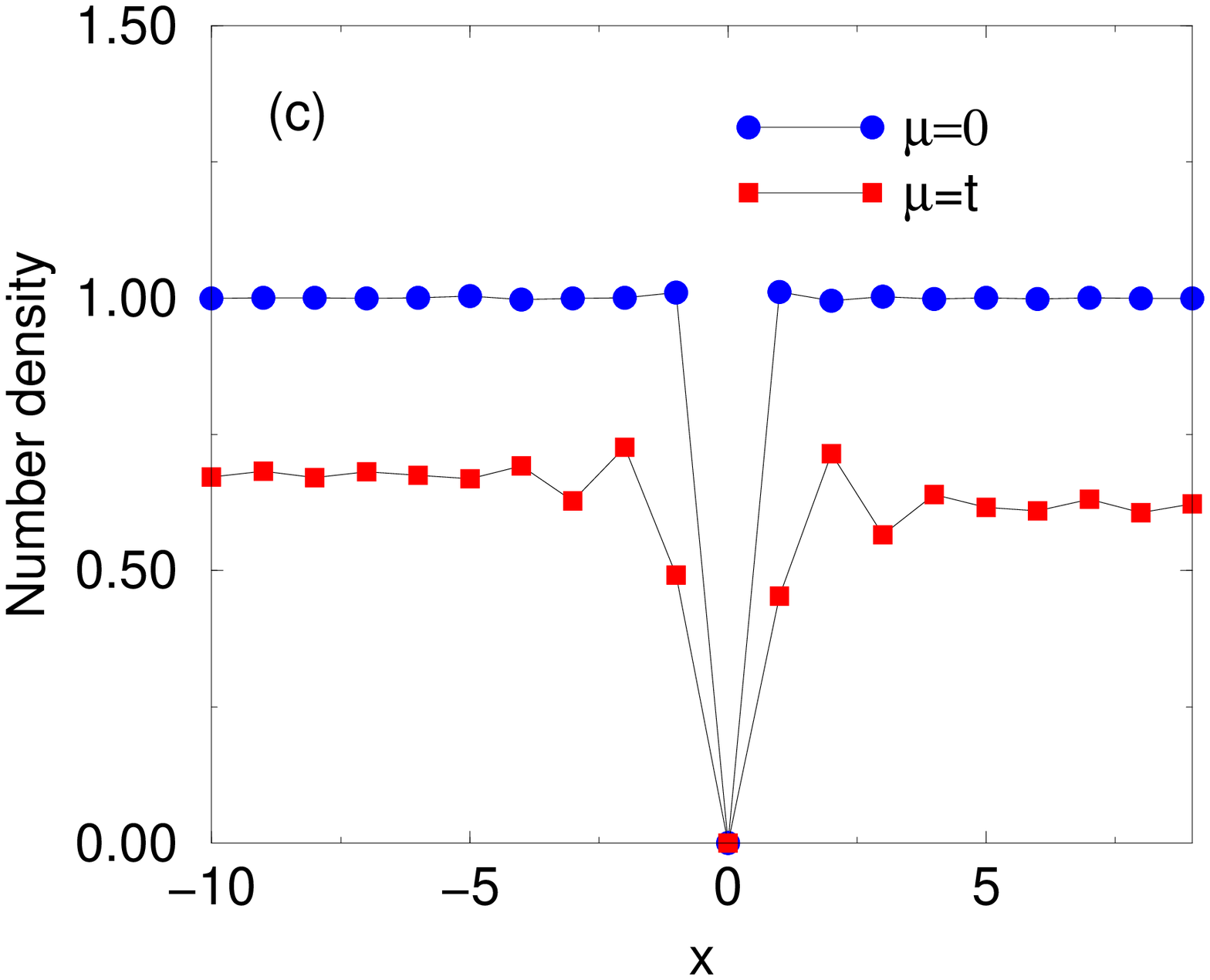,width=8.5cm}}}
\end{center}
\caption{
(a) Spatial distribution of impurities, indicated as solid
circles, corresponding to
a $d-I-n$ interface along the $[110]$ surface.
Also the labeling of the sites along the $x$ direction is shown.
(b) The magnitude of the $d$-wave component $\Delta_d$ and 
the extended $s$-wave component ($\Delta_s^{ext}$) of the
superconducting order parameter as a function of $x$, for a $d-I-n$
interface for $\mu=0,t$.
The order parameter is calculated
along the thick dashed line in direction $x$ shown in (a).
(c) The number density $n_i$
as a function of $x$ shown in (a), for a $s-I-n$
interface for $\mu=0,t$.
}  
\label{dinop}
\end{figure}

We consider then the unconventional $d$-wave 
superconductor-insulator-normal metal ($d-I-n$) interface 
shown in Fig. \ref{dinop}(a)
where $V_1=-2.5t$, and 
$\mu^{110}=100t$. We plot in Fig. \ref{dinop}(b) the magnitude of the 
$d$-wave component $\Delta_d$ and the magnitude 
of the extended $s$-wave component $\Delta_s^{ext}$ of the  
order parameter
as a function of the direction $x$ that is shown in Fig. \ref{dinop}(a),
for two different values of the chemical potential $\mu=0$ and $\mu=t$. 
It is seen that for $\mu=0$, 
$\Delta_s^{ext}$ is not modified by the presence of the interface.
In contrast $\Delta_d$ drops to zero at the interface, and increases
monotonically into a few lattice sites to the bulk value. 
For $\mu=t$ the $\Delta_d$ is much more suppressed close to the interface 
while the induced $\Delta_s^{ext}$-wave component is more enhanced.
Due to symmetry 
arguments the relation $\Delta_{\hat{x}}(r_i)=\Delta_{\hat{y}}(r_i)$
holds for the 
pairing potentials and results into a monotonic variation in the 
$d$-wave order parameter near the surface. 
This is different 
to the $s-I-n$ case, and also to the $[100]$ surface of the $d$-wave 
superconductor seen in Fig. \ref{bulkopd} 
where this increase is non-monotonous. In the 
latter case due to the translational invariance in the $y$-direction 
$\Delta_{-\hat{y}}(r_i)=\Delta_{\hat{y}}(r_i)$ is satisfied. 
The spatial oscillations are due to the atomic scale 
oscillation of the $\Delta_{\hat{x}}(r_i),\Delta_{-\hat{x}}(r_i)$ that are 
completely neglected in the quasiclassical approximation.  
We also plot in Fig. \ref{dinop}(c) the number density $n_i$ for the electrons
for the two different values of the chemical potential,
i.e., $\mu=0$ and $\mu=t$. It is similar to the case of the $s-I-n$-interface 
seen in Fig. \ref{sinop}(c)

\begin{figure}
\begin{center} 
\leavevmode 
\centerline{
\psfig{figure=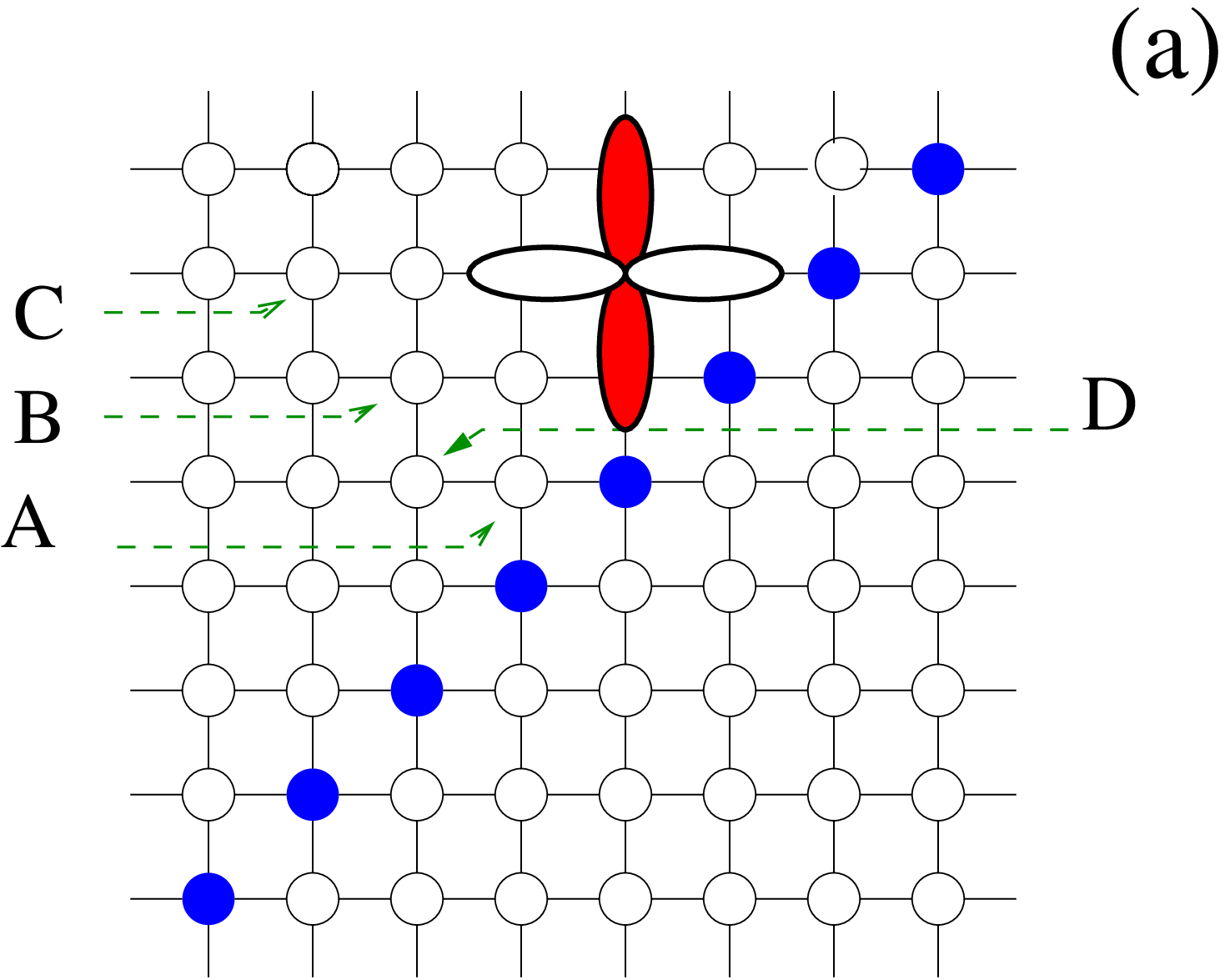,width=5.5cm,angle=0}}
\centerline{\hbox{
\psfig{figure=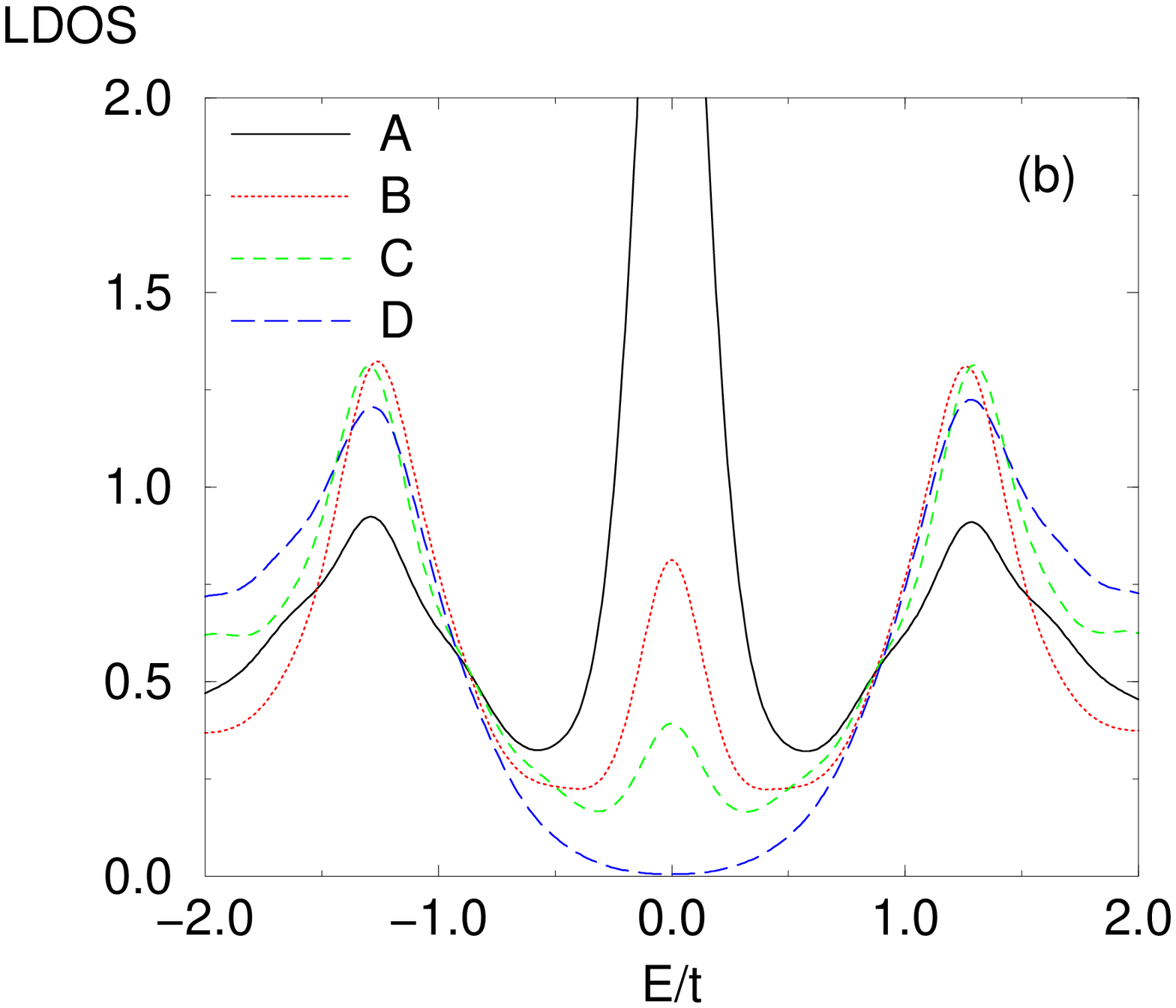,width=8.5cm}}}
\centerline{\hbox{
\psfig{figure=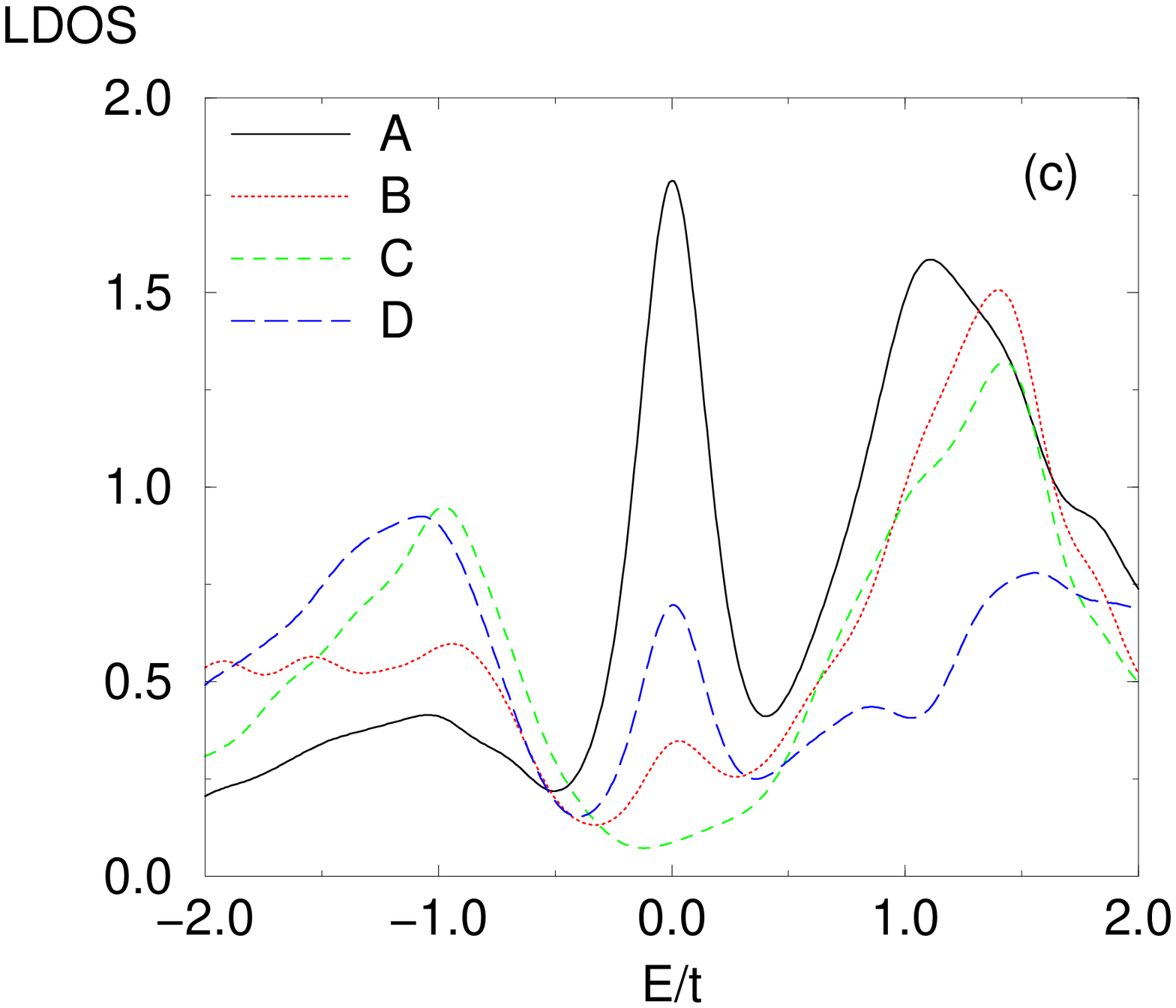,width=8.5cm}}}
\end{center}
\caption{
(a) Spatial distribution of impurities, indicated as solid
circles, corresponding to
a $d-I-n$ interface along the $[110]$ surface.
(b) The local density of states (LDOS) at $A,B,C,D$, shown in (a)
of a $d-I-n$ interface for $\mu=0$.
(c) The local density of states (LDOS) at $A,B,C,D$, shown in (a)
of a $d-I-n$ interface for $\mu=t$.
}  
\label{dindos}
\end{figure}

The main difference between the two symmetries appear in the LDOS
as we can see in Fig. ~\ref{dindos}(b) 
for the sites 
$A, B, C, D$, shown in Fig. \ref{dindos}(a), for $\mu=0$.
As expected the LDOS line shape is symmetric since $\mu=0$.
The ZBCP is formed 
and its height decreases exponentially as we move to the interior 
of the lattice along the direction perpendicular to the interface. 
However at site 
D no ZBCP is formed. 
The disappearance of ZBCP at D denotes that 
the zero energy states (ZES) wave functions have spatial variation close to the 
interface with nodes at specific sites. In this case site D 
corresponds to a node and therefore the ZBCP disappears. 
The
ZBCP is explained in the context of zero energy states 
\cite{hu} formed near the $[110]$ surfaces of $d$-wave superconductors due 
to the sign change that the quasiparticles experience in different directions 
in $k$ space. 
However the absence of ZBCP for the $[110]$ at specific 
sites is not predicted by the quasiclassical theory.
For finite $\mu=1$ seen in Fig. \ref{dindos}(c) the LDOS keeps its
$V$-like line shape. However it becomes asymmetric due to the
breakdown of the electron-hole symmetry. Also the ZBCP is reduced.
We conclude that the order parameter as well as the 
local density of states are influenced by the orientation of the interface
\cite{annett2}. 

\section{result for the surface roughness}
\begin{figure}
\begin{center} 
\leavevmode 
\centerline{
\psfig{figure=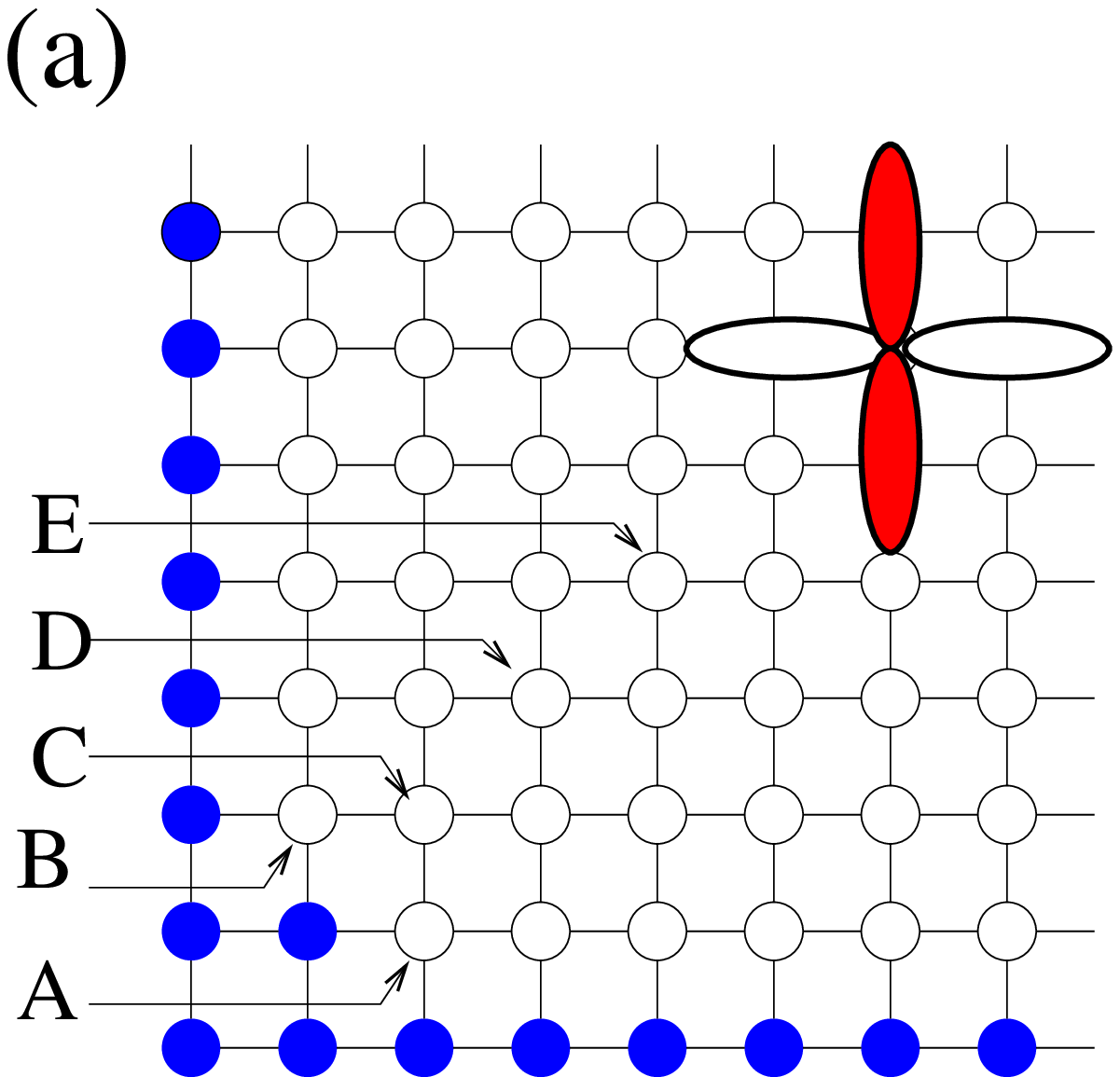,width=4.5cm,angle=0}}
\centerline{
\psfig{figure=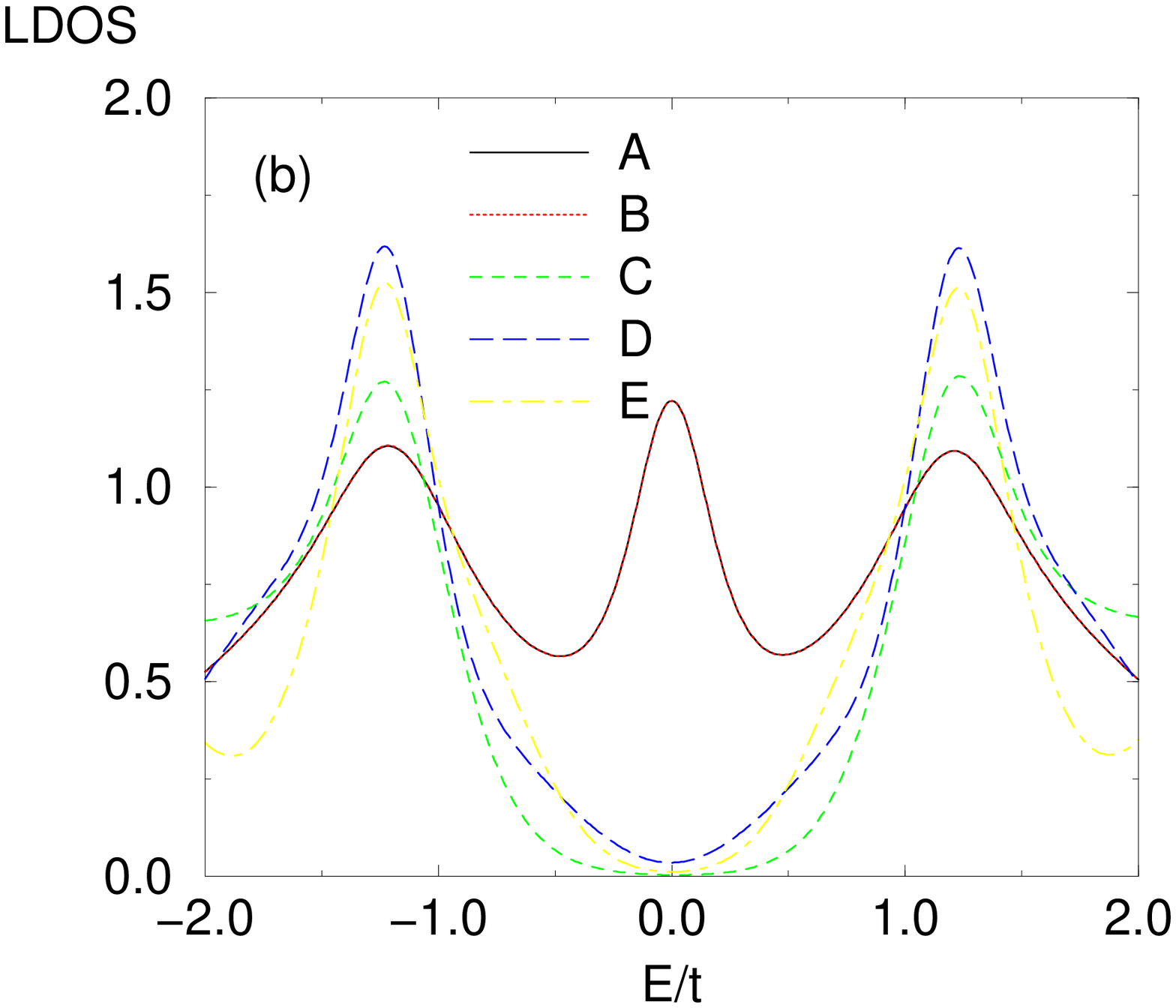,width=8.5cm}}
\centerline{
\psfig{figure=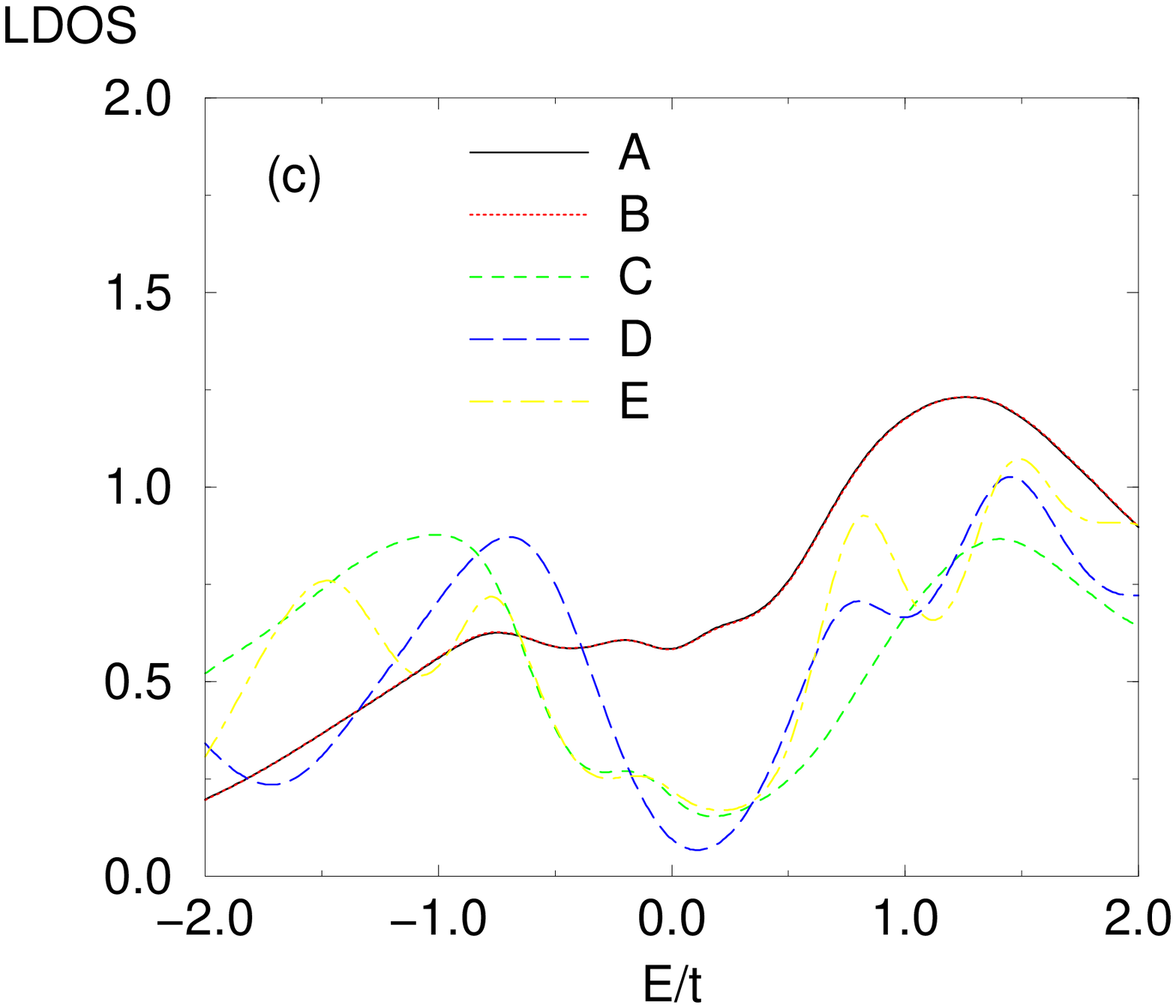,width=8.5cm}}
\end{center}
\caption{
(a) Spatial distribution of impurities corresponding 
to the corner surface, with a 
$1\times 1$ step structure. The chemical potential 
is set to $\mu^I=100t$ at the shaded surface sites.
(b) The local density of states at the specified sites
$A, B, C, D, E$ shown in (a) for $\mu=0$.
(c) The local density of states at the specified sites
$A, B, C, D, E$ shown in (a) for $\mu=t$.
} 
\label{corner}
\end{figure}

\begin{figure}
\begin{center} 
\leavevmode 
\centerline{
\psfig{figure=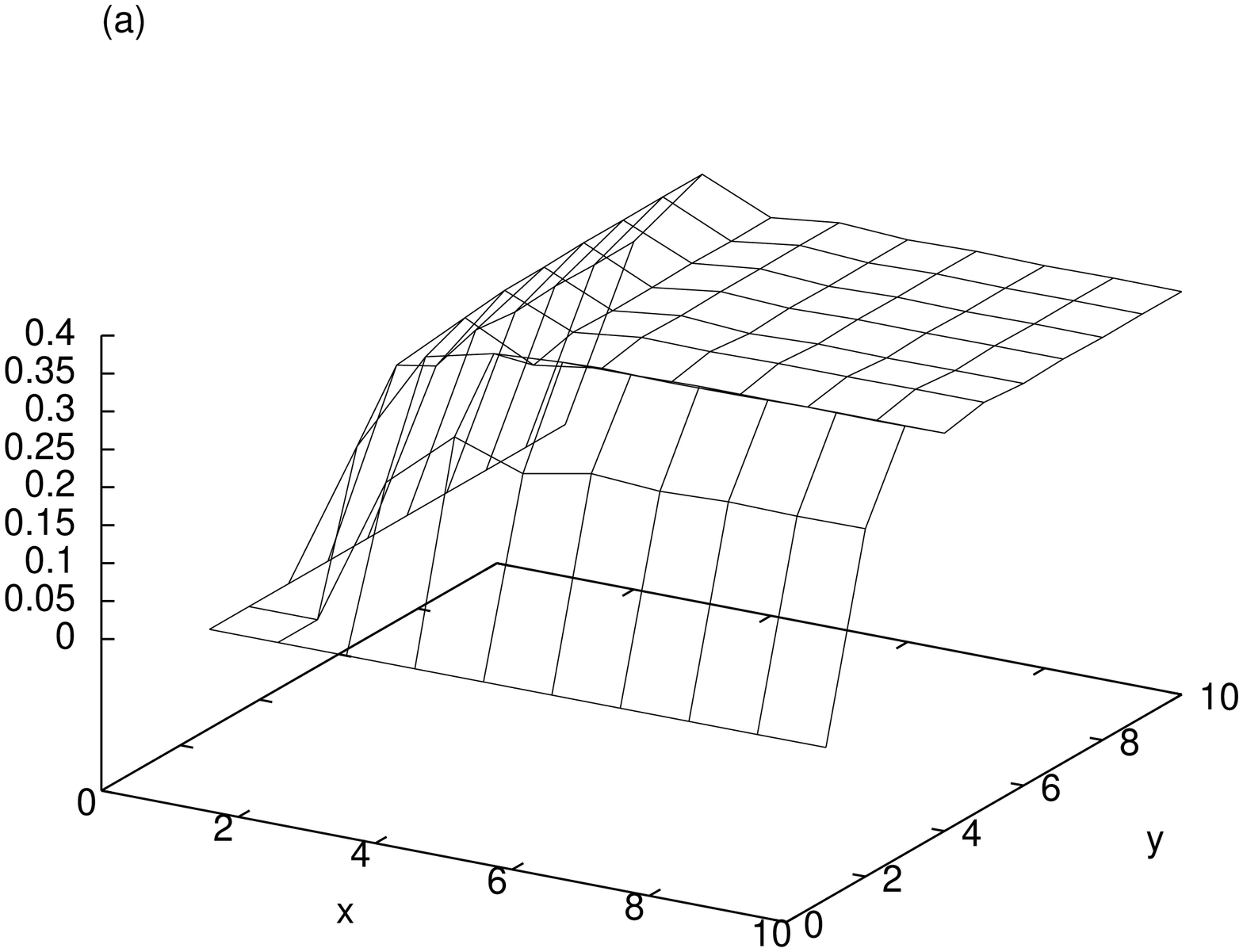,width=8.5cm,angle=-90}
            } 
\centerline{
\psfig{figure=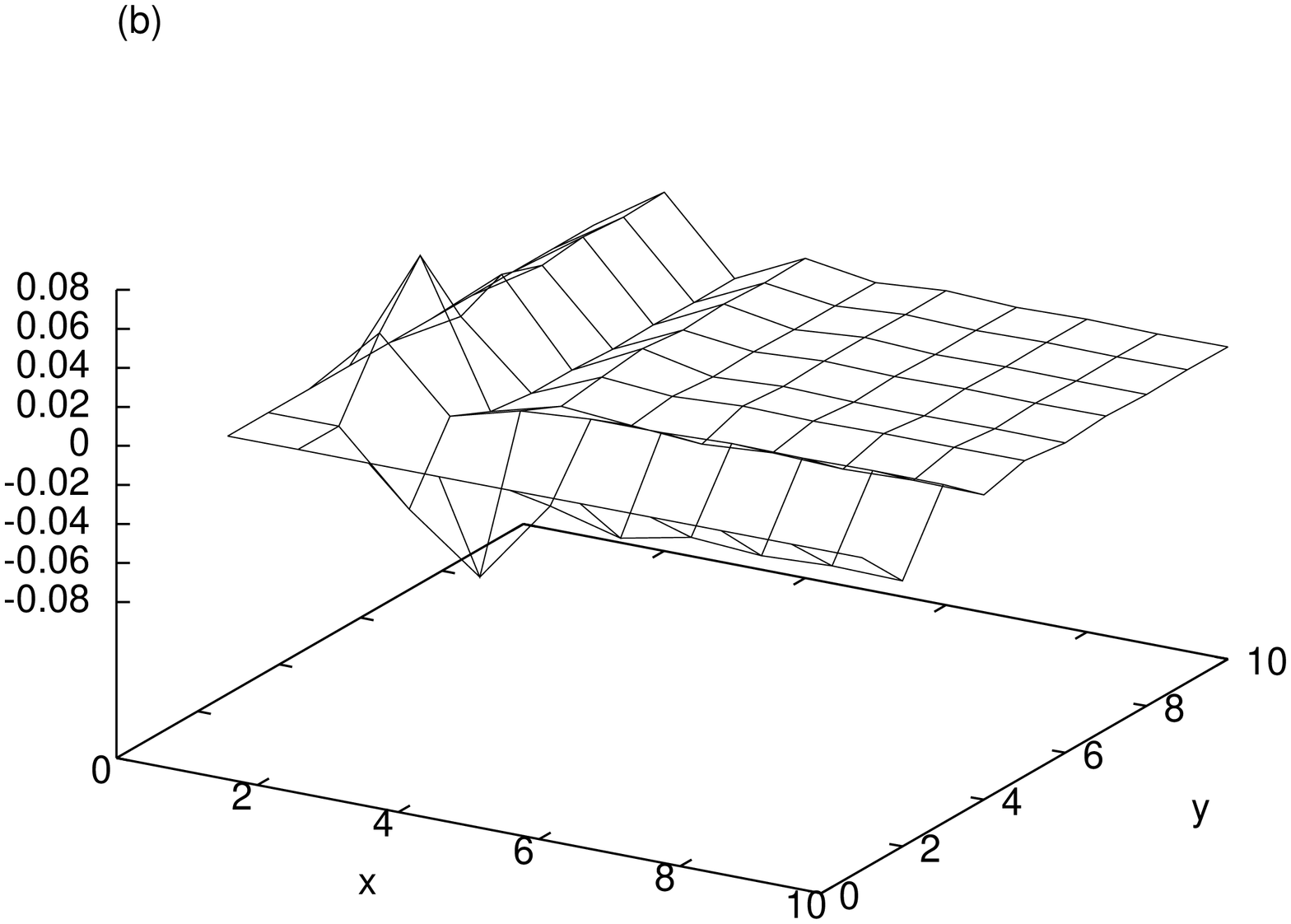,width=8.5cm,angle=-90}
}
\end{center}
\caption{Spatial dependence of the (a) $d$-wave and (b) the extended $s$-wave 
order parameter for the specified geometry of Fig. 14(a).} 
\label{cornerop}
\end{figure}

In the following we describe the effect of the surface roughness 
near the corner of the lattice. For the case of one step structure,
for $\mu=0$,
shown in Fig. \ref{corner}(a) the LDOS shows ZBCP at points 
A,B but not in C as presented in Fig. \ref{corner}(b).
Moreover the ZBCP is suppressed compared to the case of
flat $[110]$ surface seen in Fig. ~\ref{dindos}.
The quasiclassical theory predicts that for that direction
the ZBCP is maximum \cite{stefan1} since
the one step structure corresponds to the $a=\pi/4$ (where 
$a$ is the orientation of the surface).
The suppression of the ZBCP and also the disappearance of ZBCP 
from specific sites is explained 
by the spatial variation of the ZES.  
It is seen that the wave functions of the ZES form standing waves 
that decay in the bulk.
The sites A and B, that show ZBCP, correspond to an antinode 
while for the rest of the lattice sites 
the amplitude of the ZES is zero. For finite $\mu$ the ZBCP 
disappears from sites A and B due to the destructive interference 
of the ZES as shown in Fig. \ref{corner}(c). In addition the overall line 
shape of the conductance curve is asymmetric due to the breaking 
of the electron-hole symmetry.
In Fig. \ref{cornerop} the spatial variation 
of the $d$-wave and extended $s$-wave 
order parameter is plotted for $\mu=0$ at sites close 
to the lattice corner. It is seen that the $d$-wave order parameter 
is suppressed at the impurity site while the extended $s$-wave 
order parameter is not much influenced.

\begin{figure}
\begin{center} 
\leavevmode 
\centerline{
\psfig{figure=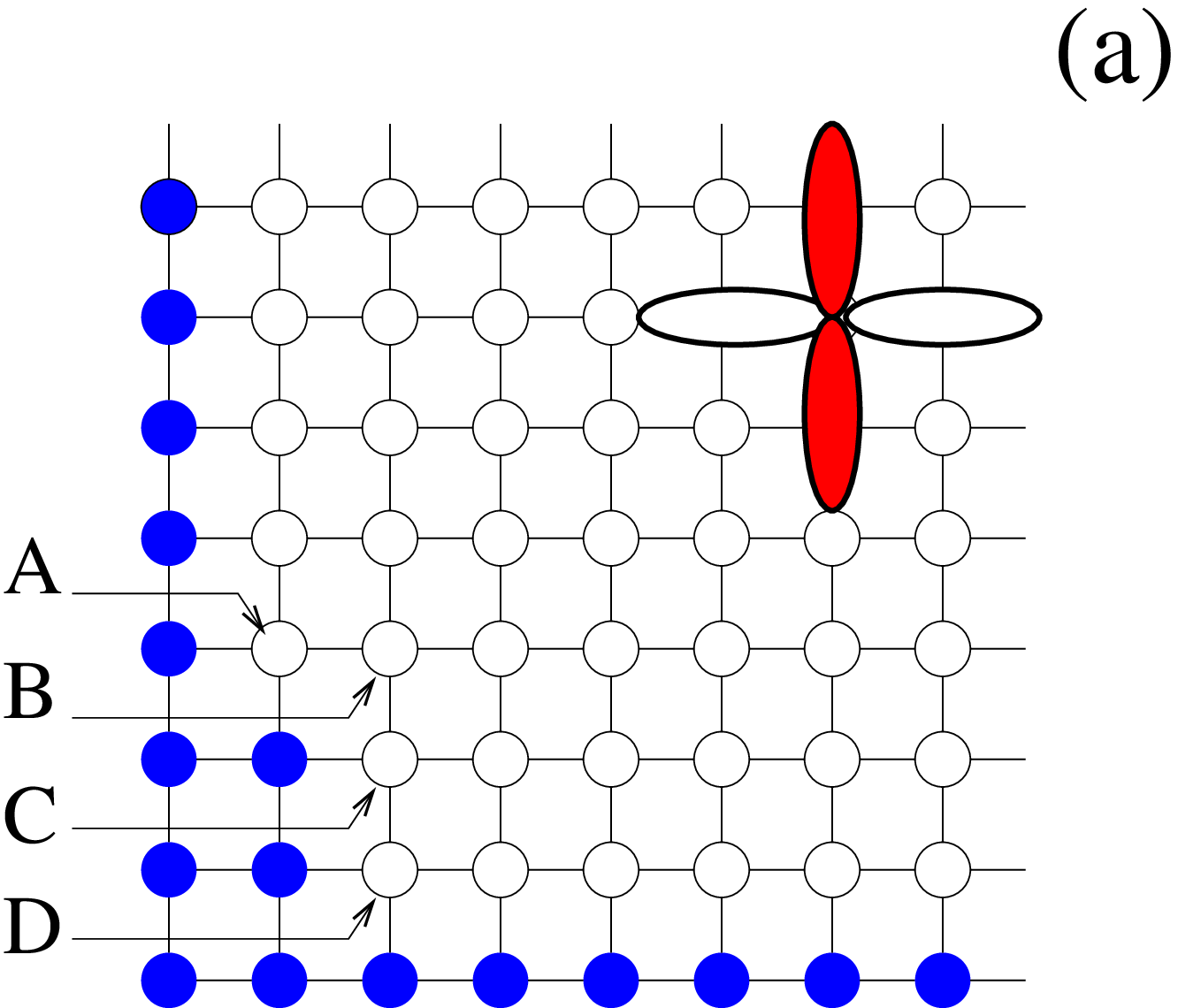,width=4.5cm,angle=0}}
\centerline{
\psfig{figure=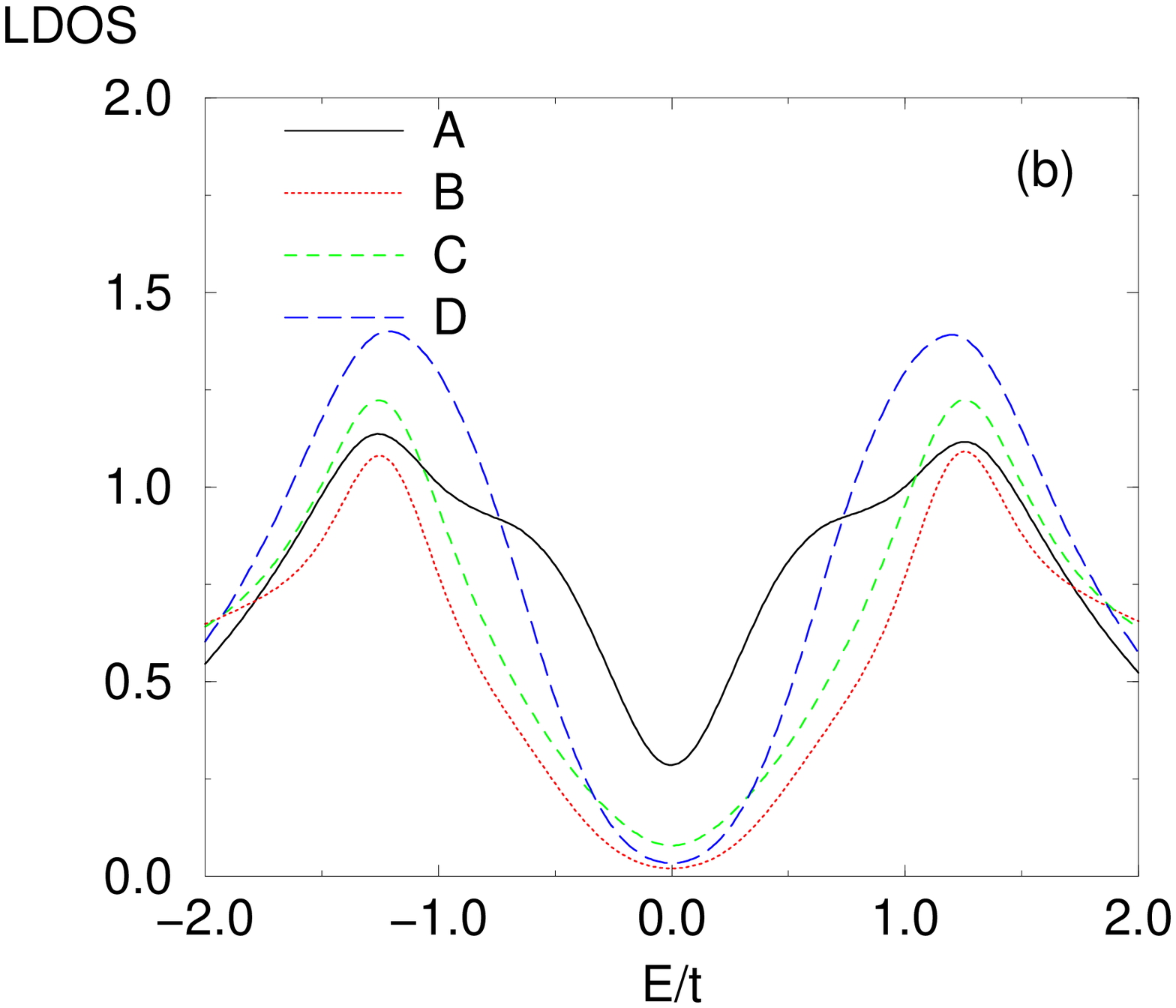,width=8.5cm}}
\centerline{
\psfig{figure=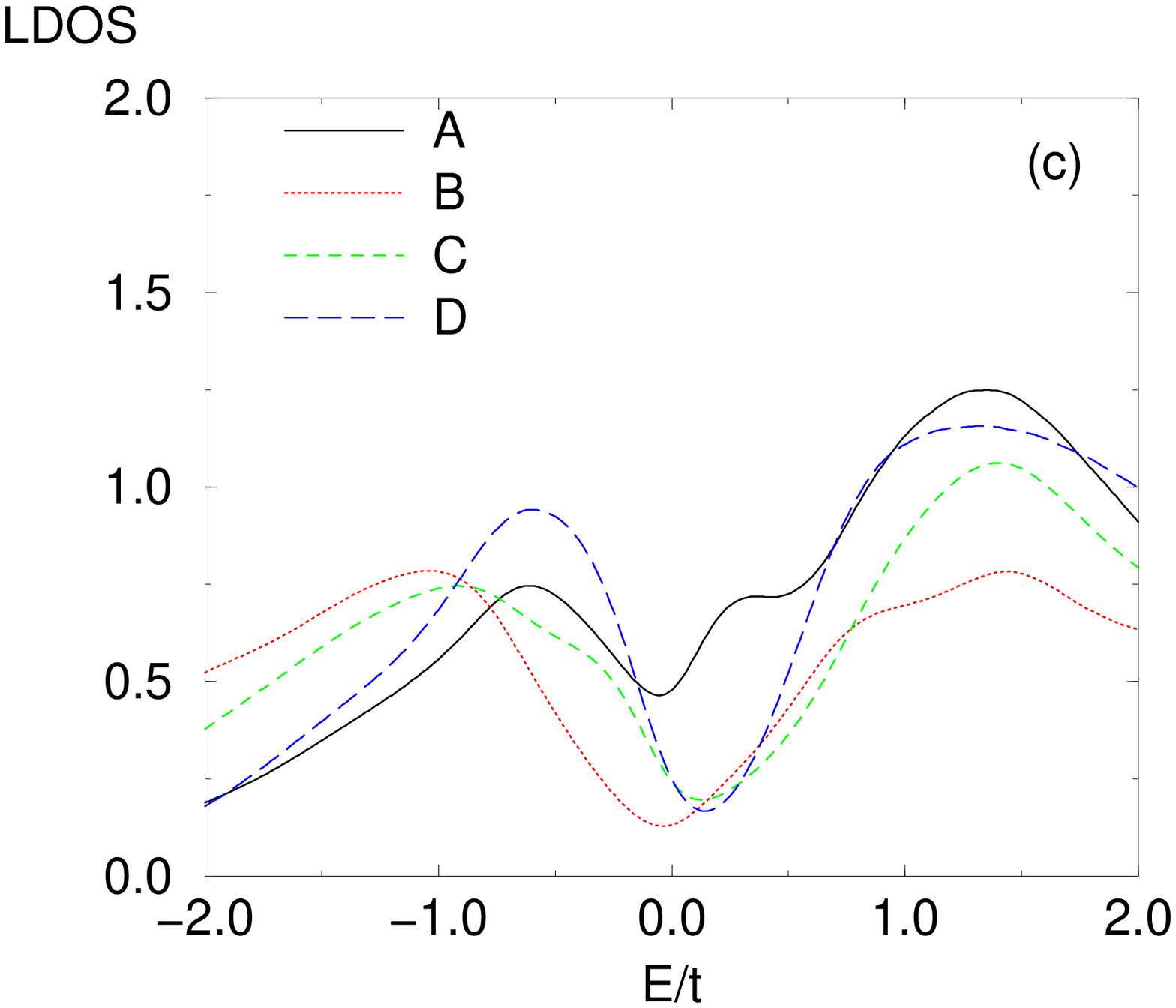,width=8.5cm}}
\end{center}
\caption{
(a) Spatial distribution of impurities corresponding 
to the corner surface, with a 
$1\times 2$ step structure. The chemical potential 
is set to $\mu^I=100t$ at the shaded surface sites.
(b) The local density of states at the specified sites
$A, B, C, D$ shown in (a) for $\mu=0$.
(c) The local density of states at the specified sites
$A, B, C, D$ shown in (a) for $\mu=t$.
} 
\label{corner3}
\end{figure}

For the $1\times 2$ step structure shown in Fig. \ref{corner3}(a)
the LDOS, presented in Fig. \ref{corner3}(b) for $\mu=0$, at points 
A,B,C,D shows no ZBCP. The quasiclassical theory predicts a ZBCP 
since this geometry corresponds to a surface tilted from $a=0$ or $a=\pi /2$.
The absence of ZBCP is explained by the destructive interference
of the standing waves and also by the asymmetricity of the 
structure. For $\mu=t$ shown in Fig. \ref{corner3}(c) some tiny 
conductance peak recovers.
 
\begin{figure}
\begin{center} 
\leavevmode 
\centerline{
\psfig{figure=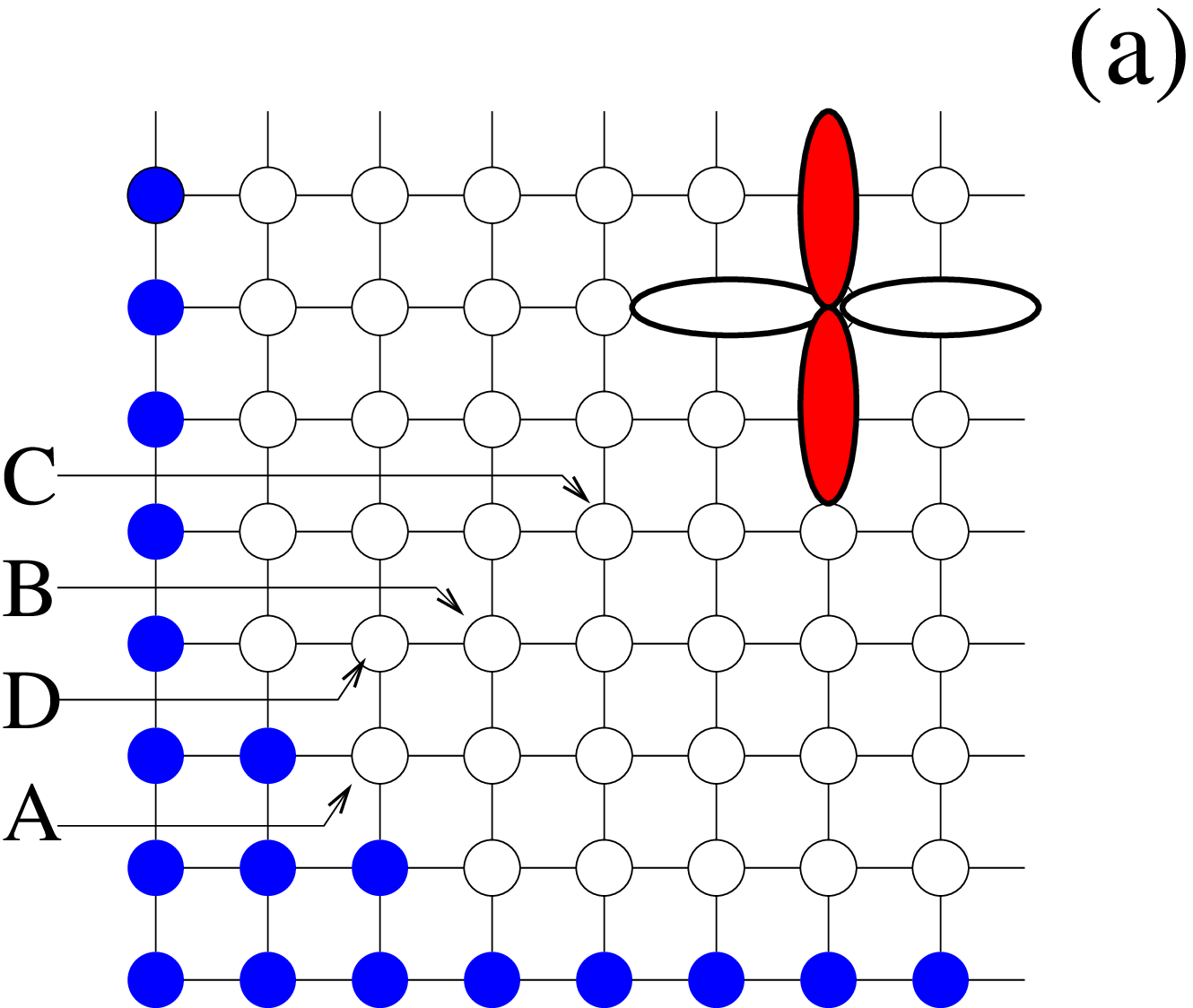,width=4.5cm,angle=0}}
\centerline{
\psfig{figure=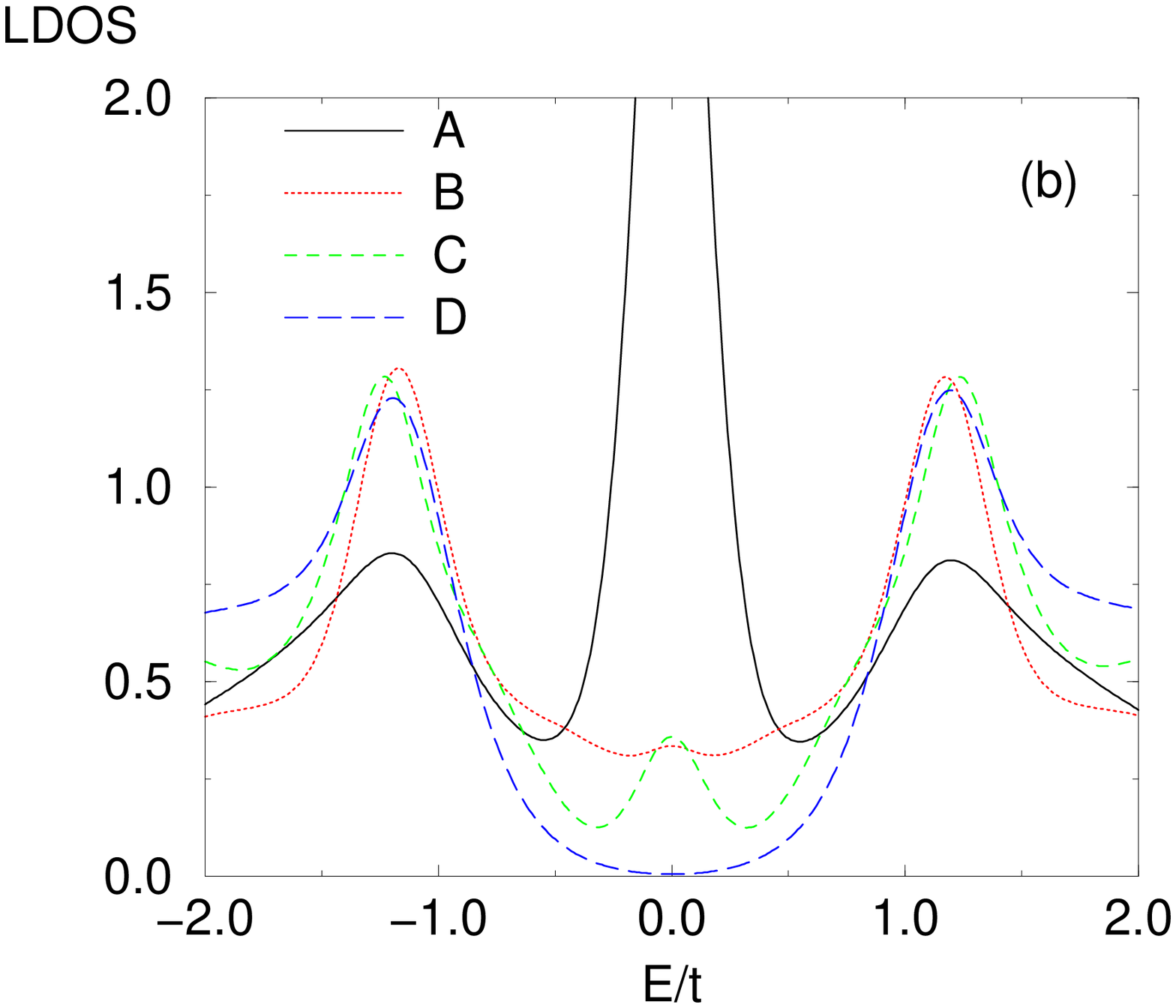,width=8.5cm}}
\centerline{
\psfig{figure=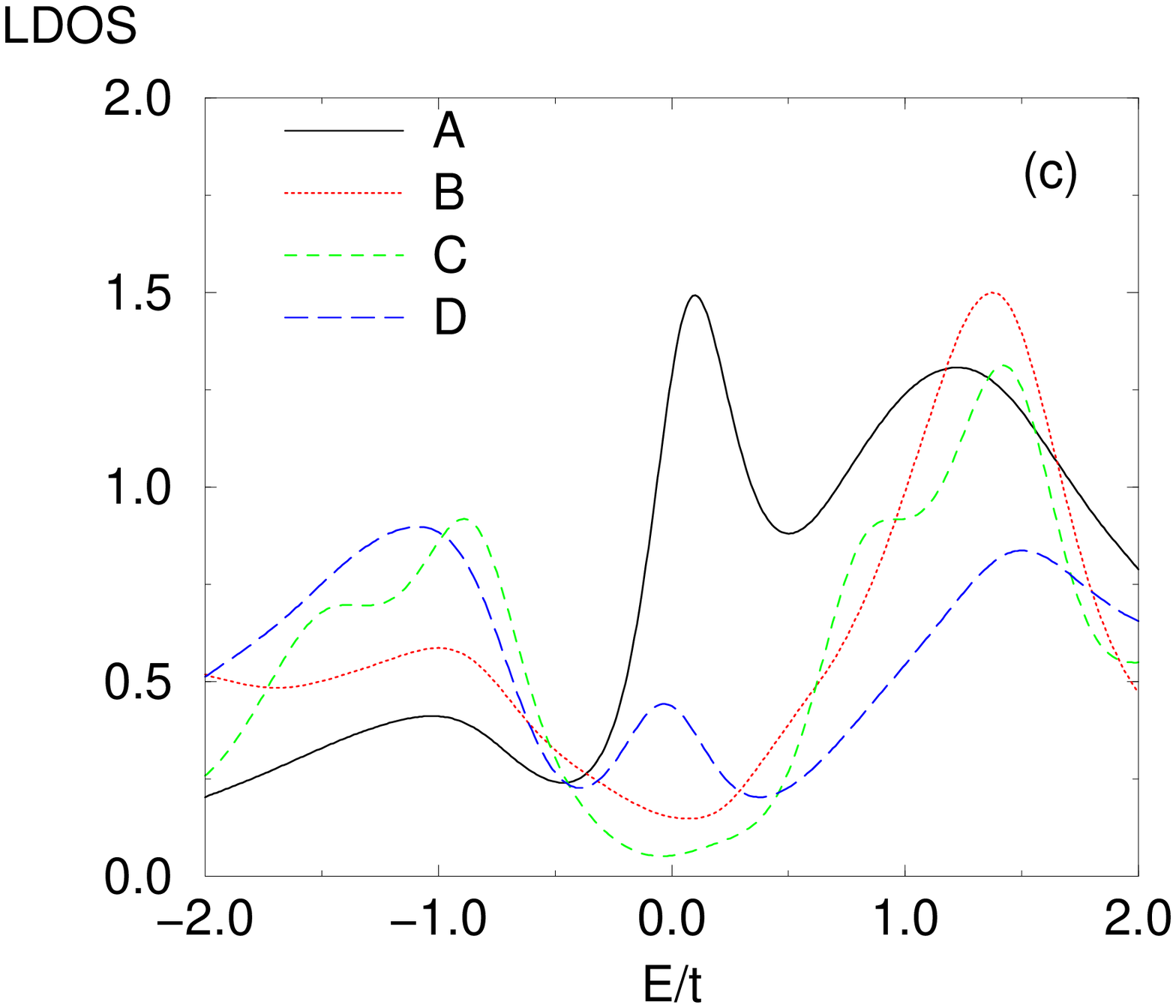,width=8.5cm}}
\end{center}
\caption{
(a) Spatial distribution of impurities corresponding 
to the corner surface, with a 
$[110]$ structure. The chemical potential 
is set to $\mu^I=100t$ at the shaded surface sites.
(b) The local density of states at the specified sites
$A, B, C, D$ shown in (a) for $\mu=0$.
(c) The local density of states at the specified sites
$A, B, C, D$ shown in (a) for $\mu=t$.
} 
\label{corner2}
\end{figure}

The ZBCP at the topmost sites recovers for the geometry shown in 
Fig. \ref{corner2}(a) for sites A,C, for $\mu=0$ as presented 
in Fig. \ref{corner2}(b). However unlike the flat $[110]$ surface no 
ZBCP is formed in site B due to the destructive interference of 
the standing waves. Moreover when finite chemical potential 
is introduced as shown in Fig. \ref{corner2}(c) the spatial distribution 
of the ZES is disturbed and for example the ZBCP appears even for 
sites, e.g., D, where normally for $\mu=0$ is absent.

\section{conclusions}
We calculated the LDOS and the order parameter of a two dimensional 
lattice of $d$-wave superconductor within the extended Hubbard 
model, self consistently. 
The dominant order parameter decays monotonically for $[110]$ interface and 
$d$-wave order parameter symmetry, while non-monotonically 
for $[100]$ interface, $d$-wave or $s$-wave order parameter symmetry. 
The induced extended $s$-wave that decays to zero in the bulk 
changes sign at the topmost sites at either side of the lattice similarly 
to the case near impurities, and twin boundaries.
The LDOS is symmetric when $\mu=0$ and it
becomes asymmetric when $\mu$ deviates from zero.
The presence of surface roughness at the corner strongly modifies 
the quasiparticle properties near the corner. The ZBCP, which is 
absent for perfect corner consistently to the quasiclassical theory, 
appears when the roughness at atomic size is introduced due to the 
oscillatory form of the bound states that decay in the bulk.
The last result is beyond the quasiclassical approximation.
The sensitivity of the properties on the atomic scale has to 
be taken into account for the correct interpretation of the 
experiments on corner junctions and in models that calculate 
the Josephson current phase relation, or the critical currents in those 
junctions taking into account the Andreev bound states.

\bibliographystyle{prsty}

\end{document}